\documentclass[pra,reprint,superscriptaddress,frontmatterverbose,aps,prstper]{revtex4-2}

\usepackage{graphicx}
\usepackage{dcolumn}
\usepackage{bm}
\usepackage{comment}
\usepackage{placeins}
\usepackage{svg}
\usepackage{booktabs}
\usepackage{tabularx}
\usepackage{subcaption}
\usepackage[breaklinks]{hyperref}  
\hypersetup{colorlinks=true,urlcolor=blue,citecolor=blue,linkcolor=blue}   
\usepackage{enumitem}          
\setlist{nosep}                 

\PassOptionsToPackage{hyphens}{url}\usepackage{hyperref}
\usepackage{url}

  {\list{}{\leftmargin=0.3in\rightmargin=0.3in}\item[]}%
  {\endlist}

\begin{document}

\title{Global Blind Spot in Understanding Trigonometric Derivatives: \\A Multinational Analysis}

\author{Yossi Ben-Zion}\email{benzioy@biu.ac.il}\affiliation{Department of Physics, Bar Ilan University, Ramat Gan 52900, Israel}
\author{Jianqiang Liu}\affiliation{School of Physics, National Physical Experiment Teaching Demonstration Center, Shandong University, Jinan 250100, People’s Republic of China}
\author{Chun-Ming Wang}\affiliation{School of Physics, Shandong University, Jinan 250100, People’s Republic of China}
\author{Atanu Rajak}\affiliation{Department of Physics, Indian Institute of Technology, Hyderabad 502284, India}
\author{Noah D. Finkelstein}\affiliation{Department of Physics, University of Colorado Boulder, Boulder, Colorado 80309, USA}

\date{\today}

\begin{abstract}
Trigonometric derivatives are fundamental in both mathematics and physics, yet their proper application, particularly the distinction between radians and degrees, poses a significant challenge for college students globally. This study identifies a widespread "blind spot" in understanding trigonometric derivatives and their implications for physical systems, highlighting a critical gap in physics education. A multinational survey of 769 college students, primarily undergraduate and graduate STEM majors, from Israel, the United States, China, and India assessed their ability to differentiate between radians and degrees in mathematical and physical contexts, focusing on harmonic motion. Results reveal that only 26.3\% of students correctly identified that the well-known expressions for trigonometric derivatives hold exclusively in radians, while 70.7\% incorrectly assumed both radians and degrees are valid. Notably, students demonstrated improved recognition of radians in physical contexts (59.0\% correct responses) compared to mathematical ones, suggesting that students rely on familiar physical equations as cognitive reference points when applying mathematical concepts. These misunderstandings appear worldwide, suggesting a universal challenge. The findings highlight the need for curriculum reforms to better connect mathematical formalism with physical application.

\end{abstract}


\maketitle

\pagestyle{plain}
\thispagestyle{plain}

\section{Introduction}
\begin{quote}
    ``An alien culture might not use 360 at all. On the other hand, I believe 2$\pi$ would be discovered in any advanced civilization.''  
    \hfill -- R. Shankar, \textit{Fundamentals of Physics I} \cite{shankar2019fundamentals}
\end{quote}

Shankar’s remark highlights a fundamental reality in mathematical physics: radians are not an arbitrary choice but a natural consequence of how we define and manipulate angular quantities. Despite their fundamental role in calculus and physics, students often fail to grasp why radians, rather than degrees, form the correct mathematical framework for differentiation and physical equations. This confusion extends beyond simple unit conversion; it likely reflects a deeper conceptual gap in the way students bridge mathematical formalism with physical intuition. In this study, we investigate how this misunderstanding manifests across diverse educational backgrounds and international contexts, and examine its implications for learning and applying fundamental principles in physics.

The successful learning of physics fundamentally depends on students' ability to apply mathematical concepts in physical contexts. This connection between mathematical understanding and physics comprehension becomes particularly critical when dealing with trigonometric functions and their derivatives. Recent research in physics education has highlighted persistent challenges students face when transferring mathematical knowledge to physical applications \cite{PhysRevPhysEducRes.12.020134,10.1119/1.17077,PhysRevSTPER.11.010108,PhysRevSTPER.11.020129,PhysRevSTPER.11.020131,PhysRevSTPER.7.010113,PhysRevSTPER.8.010111,10.1063/1.2177017,PhysRevPhysEducRes.16.010111,PhysRevPhysEducRes.16.020121,PhysRevPhysEducRes.17.010138}.

Research also indicates that trigonometry poses significant challenges for physics students at all levels, from beginners to advanced learners. Galle and Meredith \cite{10.1119/1.4862119}, in their study of life science students, identified a common difficulty in understanding the physical applications of trigonometric functions. They highlighted students’ struggles to connect the mathematical representation of trigonometric functions with their physical manifestations, such as harmonic motion or waves. Additionally, students often found it challenging to interpret angular measurements quantitatively, especially when transitioning between graphical representations and algebraic solutions.

Breitenberger \cite{weber2005students} further demonstrated how these challenges persist among advanced physics students. His work documented how, despite procedural fluency in basic mathematical analyses, many students fail to grasp the deeper conceptual meaning of trigonometric functions. For instance, students tend to perceive these functions merely as computational tools, without understanding the relationship between their geometric properties and physical applications. This leaves a substantial gap between theoretical comprehension of trigonometry and its use in solving complex physics problems.

While previous research has extensively documented students' difficulties with trigonometry, little attention has been given to the specific impact of angular representation— degrees versus radians— on their ability to correctly apply differentiation in physics. This study aims to fill that gap by systematically analyzing how students across diverse educational systems understand and utilize angle units in differentiation and assessing the implications for problem-solving in physics. Previous studies have focused on general difficulties with trigonometric functions and their applications, but have not sufficiently addressed the critical distinction between these units and their influence on the rate of change of functions in physical systems, such as harmonic motion and other physical systems. Bridging this gap is essential for ensuring that students can apply trigonometry effectively across a wide range of physical problems.

Angular measurements are ubiquitous in physics, from basic kinematics to quantum mechanics. These angles can be quantitatively represented using either degrees or radians, with each system having distinct historical and practical origins. The division of a circle into 360 degrees, originating from ancient Babylonian mathematics, remains deeply ingrained in everyday understanding. In contrast, the radian measure, representing the ratio of arc length to radius, emerged as a mathematically natural unit, with \(2\pi\) radians corresponding to a full circle. This duality in representation creates a fascinating interface between mathematical formalism and physical understanding.

Remarkably, our research reveals that most students exhibit fundamental gaps in both mathematical distinction and physical application of these representations. These findings transcend academic disciplines, educational levels, and most significantly, cross continental and cultural boundaries, suggesting a global blind spot that demands attention from curriculum designers and instructors in both mathematics and physics.

The everyday prevalence of degrees (e.g., "he made a 180-degree turn in his opinion") contrasts sharply with the mathematical preference for radians in advanced studies. This dichotomy is reflected in educational progression: students typically first encounter angles in degrees during early mathematics education, with radians introduced later. At the university level, particularly in calculus and trigonometry courses, radians become the exclusive representation, to the extent that degree-based trigonometric functions are rarely used in textbooks and instruction. This discontinuity in teaching different representations and the transition between them can create a fundamental confusion among students, serving as a primary motivation for the current research.

In introductory physics courses, especially at the high school level or in algebra-based college physics, there exists a hybrid approach to these representations. Early mechanics and kinematics typically employ degrees, allowing students to visualize problems using familiar units (e.g., projectile motion at 30 degrees rather than $\pi/6$ radians). However, when discussing periodic motion and using trigonometric functions to represent position as a function of time, instruction often shifts to radians. This transition is often justified with the simple explanation that radians are ``more natural'' due to their relationship to arc length, without deeper exploration of the mathematical implications. This approach is commonly found in introductory physics textbooks (e.g. \cite{shankar2019fundamentals,halliday2013fundamentals,serway2000physics,tipler2007physics})

The critical distinction lies not in the dimensionless nature of these units—both degrees and radians are dimensionless—but in their derivatives. The rate of change in radians differs from that in degrees by a factor of $\pi/180$:

\begin{equation}
\frac{d}{dx}\sin(x) = \cos(x) \text{ [in radians]}
\end{equation}

\begin{equation}
\frac{d}{dx}\sin(x^{\circ}) = \frac{\pi}{180}\cos(x^{\circ}) \text{ [in degrees]}
\end{equation}

\noindent This distinction becomes particularly crucial in the context of harmonic oscillation, where the fundamental equation of motion:

\begin{equation}
\frac{d^2x}{dt^2} = -\omega^2x
\end{equation}

\noindent relies on the radian representation for its standard solutions. The familiar expression for position as a function of time in harmonic motion, 
\begin{equation}
x(t) = A\cos (\omega t + \phi)
\end{equation}
where \( \omega = \sqrt{\frac{k}{m}} \) is valid only when using radians, a fact that remains unclear to many students throughout their academic careers (demonstrated below). This distinction becomes particularly crucial not only in harmonic oscillation but also in a wide range of physical systems where accurate mathematical representations are essential for modeling and understanding systems' behaviors.

This study investigates students' understanding of trigonometric derivatives in both mathematical and physical contexts, with particular emphasis on the distinction between degrees and radians. We examine how this understanding varies across different countries, educational systems, and academic backgrounds, while also exploring the relationship between students' confidence levels and their conceptual understanding. Through this comprehensive analysis, we aim to identify patterns in student reasoning and that may be used to inform pedagogical interventions to address this fundamental gap in physics and mathematics education.

\section{Research Questions}

The study focuses on the following primary research questions:
\begin{enumerate}
    \item To what extent do university students correctly identify the derivatives of trigonometric functions in radians versus degrees? 
    \item To what extent do students correctly identify the difference between degrees and radians in a physical system such as a harmonic oscillator?
    \item Are there significant differences in understanding these questions between students from different countries and educational systems?
     \item What is the correlation between students' confidence levels and the correctness of their answers to conceptual questions involving the trigonometric expression for harmonic oscillators?

\end{enumerate}

In addition to these primary questions, the following sub-questions will be explored:

\begin{enumerate}
    \item What patterns emerge from students' explanations regarding the validity of the trigonometric expression for a harmonic oscillator, and how do these patterns correlate with the correctness of their answers and their confidence levels?

    \item To what extent do students' academic departments (e.g., Physics, Mathematics, Engineering) influence their understanding of trigonometric derivatives, and how do these departments influence their application of trigonometric functions in physical systems like harmonic oscillators?

    \item To what extent do grades in foundational courses, such as Mechanics and Calculus, predict students' ability to correctly identify the conditions under which trigonometric derivatives are valid, and how do these grades predict their ability to apply this knowledge in physical systems?

\end{enumerate}

\section{\label{sec:level2}Methods}

\subsection{\label{survey}The Survey}
The survey included 6 main core questions: 

\textbf{Q1:} What is the derivative of sin(t) with respect to t? (Open-ended question)

\textbf{Q2:} What is the derivative of cos(t) with respect to t? (Open-ended question)

\textbf{Q3:} The answer you gave in the previous questions is correct in the case of:

\begin{itemize}
    \item Angle in radians.
    \item Both are correct.
    \item Angle in degrees.
\end{itemize}

\textbf{Q4:} An object with a mass of \(m\) performs simple harmonic motion under a spring connection with an elastic coefficient of \(k\), and its position is described as \( x(t)=Acos(\omega t+\varphi)\), where \(A\) and  \(\varphi\) are determined by the initial state, and  \( \omega = \sqrt{\frac{k}{m}} \) . 

The expression \( x(t)=Acos(\omega t+\varphi)\) is correct in the case of:
\begin{itemize}
    \item Angle in radians.
    \item Both are correct.
    \item Angle in degrees.
    
\end{itemize}

\textbf{Q5:} Please briefly explain your answer to Q4. (Open-ended question)

\textbf{Q6:} How sure are you that the answer you gave (in question Q4) is correct?
\begin{itemize}
    \item Very sure.
    \item Somewhat sure.
    \item Natural.
    \item Somewhat unsure.
    \item Not sure at all/Very Unsure.
    
\end{itemize}

In addition to these core survey questions, additional data were collected to support a more comprehensive analysis of students' reasoning patterns, confidence levels, and demographic influences. These supplementary questions were incorporated into the statistical analyses and are presented in the results section.

The questionnaire was identical as possible across all countries, with translations provided as necessary. In the United States and India, the survey was conducted in English. In Israel, it was translated into Hebrew, and in China, it was translated into Mandarin Chinese. The translations were performed by physics experts fluent in the respective languages to ensure accuracy and consistency across versions.

\subsection{Participants}

The sample consisted of 769 participants from four primary countries: Israel (35.4\%), the United States (21.3\%), China (24.2\%), and India (19.1\%). The average age of participants was 21.37 years (SD = 3.01). Regarding gender distribution, the majority identified as male (67.1\%), followed by female (31.6\%), and a small proportion identified as other (1.3\%). Most participants were pursuing a Bachelor's degree (84.2\%), while smaller proportions were pursuing Master's (7.8\%), Doctorate (6.9\%), or Postdoctoral/Other degrees (1.0\%). Participants were predominantly in their first year of study (50.3\%), with fewer in their second (28.3\%), third (14.1\%), fourth (5.9\%), fifth (1.1\%), or sixth year (0.4\%). The largest group of participants majored in physics and any other field (50.3\%), followed by engineering (21.9\%), chemistry, life sciences, or neuroscience (14.5\%), math and computer science (8.1\%), and other fields (5.3\%) (see Table \ref{tab:1}).

\begin{table}[ht]
\caption{\label{tab:1}
Demographic characteristics of the sample}
\begin{ruledtabular}
\begin{tabular}{lccccccc}
Variable & N & \% \\
\hline
\textbf{Country} & & \\
Israel & 272 & 35.4  \\
USA & 164 & 21.3 \\
China & 186 & 24.2 \\
India & 147 & 19.1 \\
\hline
\textbf{Gender} & & \\
Male & 516 & 67.1 \\
Female & 243 & 31.6 \\
Other & 10 & 1.3 \\
\hline
\textbf{Degree}		 \\		
Bachelor's & 647 & 84.2 \\
Master's & 60 & 7.8 \\
Doctorate & 53 & 6.9 \\	
Postdoctoral and Other & 8 & 1.0 \\
\hline
\textbf{Year of study} \\
Year 1 & 321 & 40.57  \\
Year 2 & 173 & 22.50 \\
Year 3 & 105 & 13.65 \\
Year 4 + & 57 & 7.41 \\
Graduate students (all years) and Other & 122 & 15.9 \\
\hline
\textbf{Major Field}\\
Physics & 386 & 50.3 \\ 
Engineering & 168 & 21.9 \\
Math and Computer Science & 62 & 8.1 \\
Chemistry, Life Sciences, or Neuroscience & 111 & 14.5 \\ 
Other & 41 & 5.3 \\
\end{tabular}
\end{ruledtabular}
\end{table}

\subsection{Data Collection Description}

The study utilized a structured questionnaire to examine students' understanding of trigonometric derivatives and their impact on physics comprehension (see Section~\ref{survey}). To minimize potential biases associated with self-reporting or online research during questionnaire completion, data were collected in class, during course lecture. 

Data collection was conducted at four leading research universities:
\begin{itemize}
    \item \textbf{United States:} University of Colorado Boulder
    \item \textbf{Israel:} Bar-Ilan University
    \item \textbf{China:} Shandong University
    \item \textbf{India:} Indian Institute of Technology (IIT) Hyderabad
\end{itemize}
These institutions were selected for their strong academic programs in science and their ability to provide a diverse and representative sample of students from various educational systems and cultural backgrounds.

\subsection{Analysis plan}
 We used SPSS version 28.0 to perform the statistical analysis. First, we produced descriptive statistics, with frequencies (N\%) for categorical variables, and means with standard deviations for continuous variables. To assess associations between two categorical variables (e.g. country vs. correctness of a question) we conduct Chi-square tests. To analyze the association between confidence level and response categories of Q3 and Q4, we conducted one way Analysis of Variance (ANOVA) of Q3 and Q4. Finally, we conducted multivariate models, by using logistic regression to predict correct response and linear regression to predict confidence interval. Statistical significance is marked at \(P<0.05\).

\section{RESULTS}

\subsection{Descriptive statistics of student success}

Table \ref{tab:2} presents the descriptive statistics of the primary dependent variables, highlighting participants' accuracy in identifying the correct case codes for trigonometric derivatives and harmonic oscillator expressions.

For trigonometric derivatives (Q3), 26.3\% of participants (202) correctly identified that the angle must be in radians for the derivative to be valid. However, 70.7\% (544) incorrectly selected the option indicating that both radians and degrees were correct, reflecting a significant misinterpretation of the conditions. A smaller fraction, 3\% (23), incorrectly assumed that the angle was in degrees.

For the harmonic oscillator expression (Q4), 59.0\% of participants (454) correctly identified the angle as being in radians. However, 34.7\% (267) again misinterpreted the conditions, selecting the option that both radians and degrees were correct. Additionally, 6.2\% (48) incorrectly assumed that the angle was in degrees.

These results underscore a widespread challenges in understanding among participants, particularly regarding the distinction between radians and degrees in both mathematical and physical contexts. The high percentage of incorrect responses in Q3 and Q4, even among students familiar with trigonometric derivatives from Q1 and Q2, highlights the "blind spot" this study aims to investigate.

\begin{table}[ht]
\caption{\label{tab:2}
Descriptive statistics of the primary dependent variables}
\begin{ruledtabular}
\begin{tabular}{lcc}
&  Correct Case Code for  & Correct Case Code for \\
     & Trig Derivatives & Harmonic Oscillator \\
     & & Expression \\\hline
Radians & 26.3\% (202) & 59.0\% (454) \\
Both & 70.7\% (544) & 34.7\% (267) \\
Degrees & 3\% (23) & 6.2\% (48) \\
\end{tabular}
\end{ruledtabular}
\end{table}

For completeness, we note that over 95\% of respondents correctly identified the derivatives of sine and cosine in Q1 and Q2. These open-ended questions allowed students to freely write their answers, revealing that all respondents used the standard radian-based expressions, with no attempts to express them in degrees. The most common errors involved misplacing the minus sign, a procedural mistake that is not the focus of this study.

\subsection{Student success by country}

\begin{figure}[ht]
\centering
\includegraphics[width=0.5\textwidth]{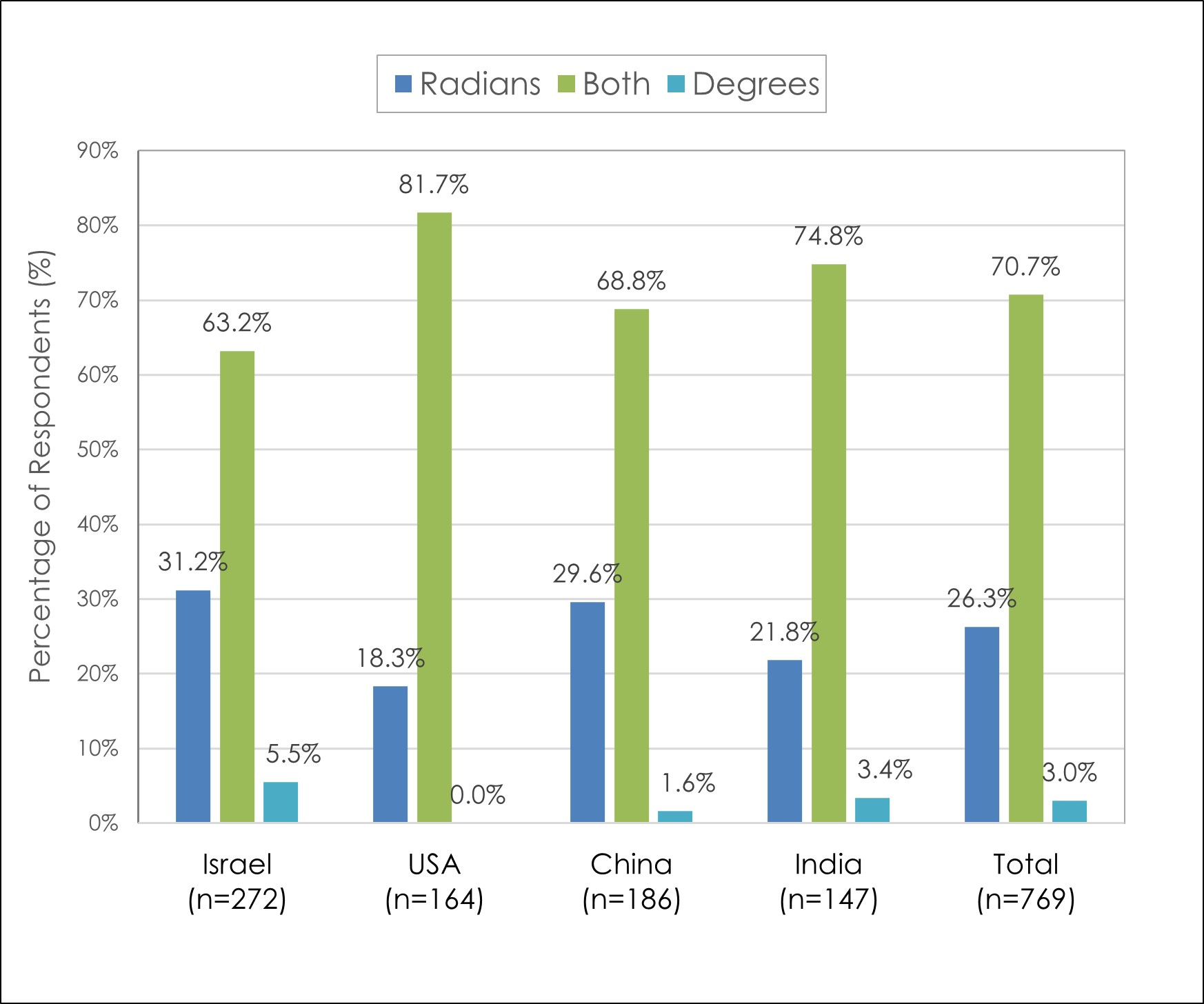}
\caption{\label{fig1:Q3_distribution} Distribution of responses for Q3 (trigonometric derivatives) across countries in percentages, highlighting the correct and incorrect responses. A detailed breakdown of percentages and absolute counts is provided in Table~\ref{tab:3} in the Appendix.}
\end{figure}

In examining the association between responses for the correct case code for trigonometric derivatives and country, a significant correlation was observed (\(\chi^2 (6) = 25.80, p < .001\)). Israel exhibited the highest rate of correct responses ("Angle in radians") at 31.3\%, followed by China (29.6\%). Lower rates were observed in the United States (18.3\%) and India (21.8\%). Regarding the incorrect response "Both are correct," Israel had the lowest rate (63.2\%), with higher rates in China (68.8\%), India (74.8\%), and the United States (81.7\%). Although a significant correlation was observed, the consistently high rates of incorrect responses across countries underscore the widespread nature of the "blind spot."  
Figure~\ref{fig1:Q3_distribution} visualizes the distribution of responses for Q3 across all countries, emphasizing the observed trends in correct and incorrect responses.
It is important to note that these results may partially reflect differences in student populations across countries, including variations in academic year and major distribution, which are not identical across the sample. A more detailed analysis of these factors is presented in later sections. Nonetheless, the consistently high rates of incorrect responses suggest that this conceptual blind spot is a global phenomenon.

\begin{figure}[ht]
\centering
\includegraphics[width=0.5\textwidth]{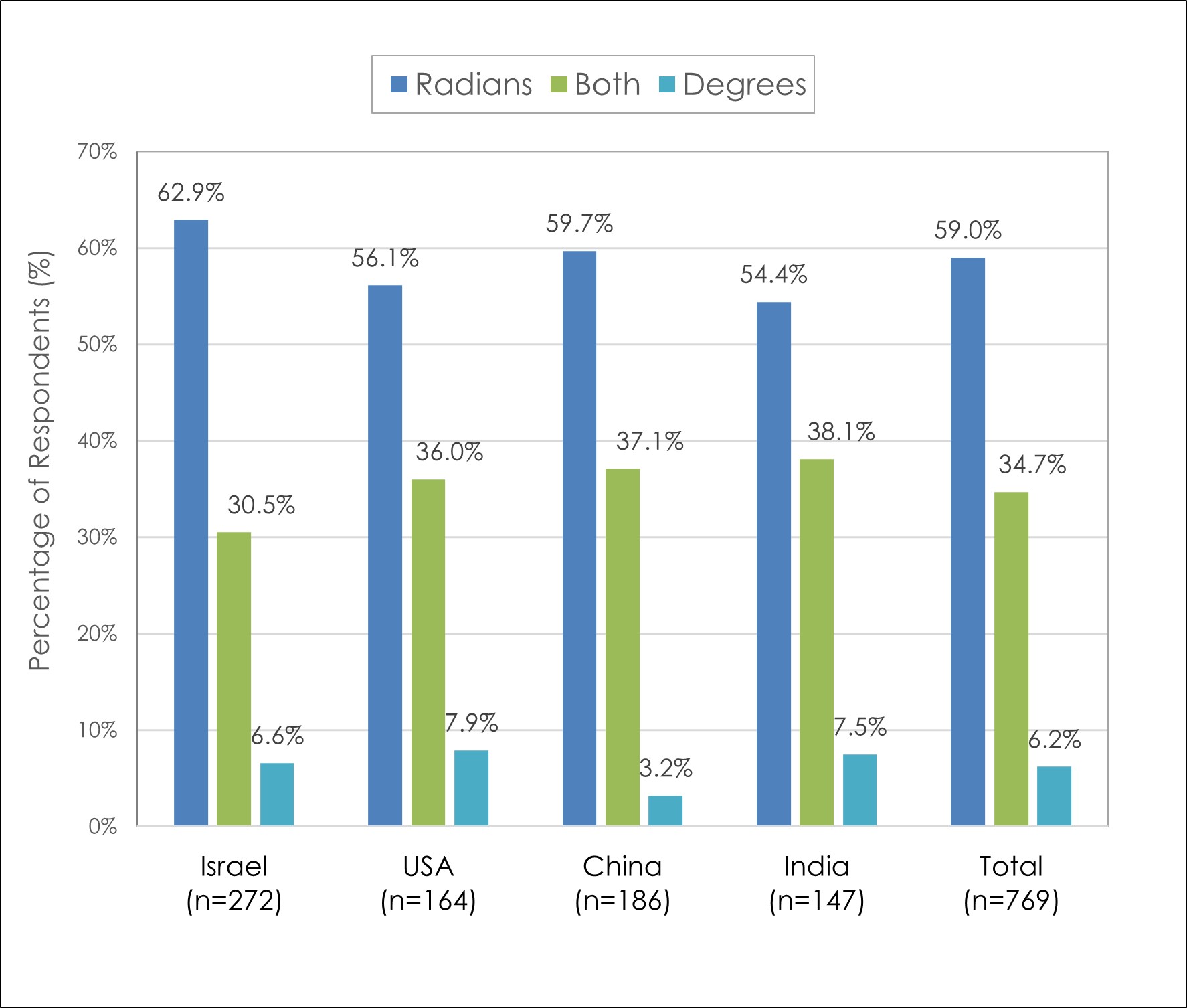}
\caption{\label{fig2:Q4_distribution} Distribution of responses for Q4 (harmonic oscillator expression) across countries in percentages, showing the observed trends in correct and incorrect responses.A detailed breakdown of percentages and absolute counts is provided in Table~\ref{tab:4} in the Appendix.}
\end{figure}

In analyzing the association between responses for the correct case code for the harmonic oscillator expression and country, no significant correlation was found (\(\chi^2 (6) = 7.58, p = .270\)). Israel exhibited the highest rate of correct responses ("Angle in radians") at 62.9\%, followed by China (59.7\%). The United States and India had slightly lower rates at 56.1\% and 54.4\%, respectively. Regarding the incorrect response "Both are correct," Israel had the lowest rate (30.5\%), with higher rates observed in the United States (36.0\%), China (37.1\%), and India (38.1\%). For the incorrect response "Angle in degrees," China had the lowest rate (3.2\%), followed by Israel (6.6\%), India (7.5\%), and the United States (7.9\%). These results reinforce the notion that the blind spot is widespread and not dependent on the country.  
Figure~\ref{fig2:Q4_distribution} illustrates the distribution of responses for Q4 across all countries, showcasing the consistent patterns in correct and incorrect answers.

\subsection{Matched Sample Analysis}

\begin{figure}[ht]
\centering
\includegraphics[width=0.5\textwidth]{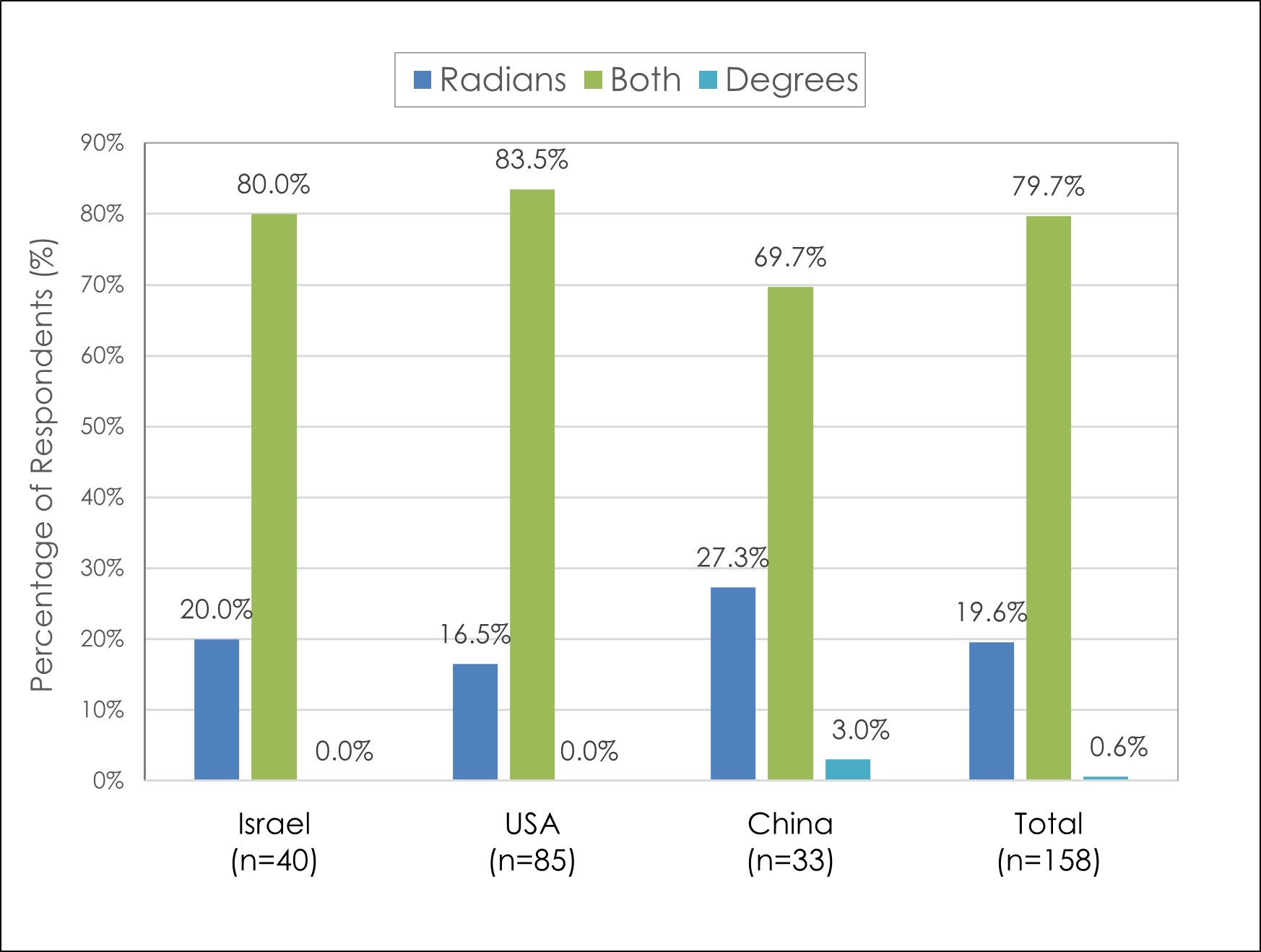}
\caption{\label{fig:matched_Q3} Distribution of responses for Q3 (trigonometric derivatives) among first- and second-year physics students, across Israel, the United States, and China. Percentages and absolute counts are provided in Table~\ref{tab:33} in the Appendix.}
\end{figure}

\begin{figure}[ht]
\centering
\includegraphics[width=0.5\textwidth]{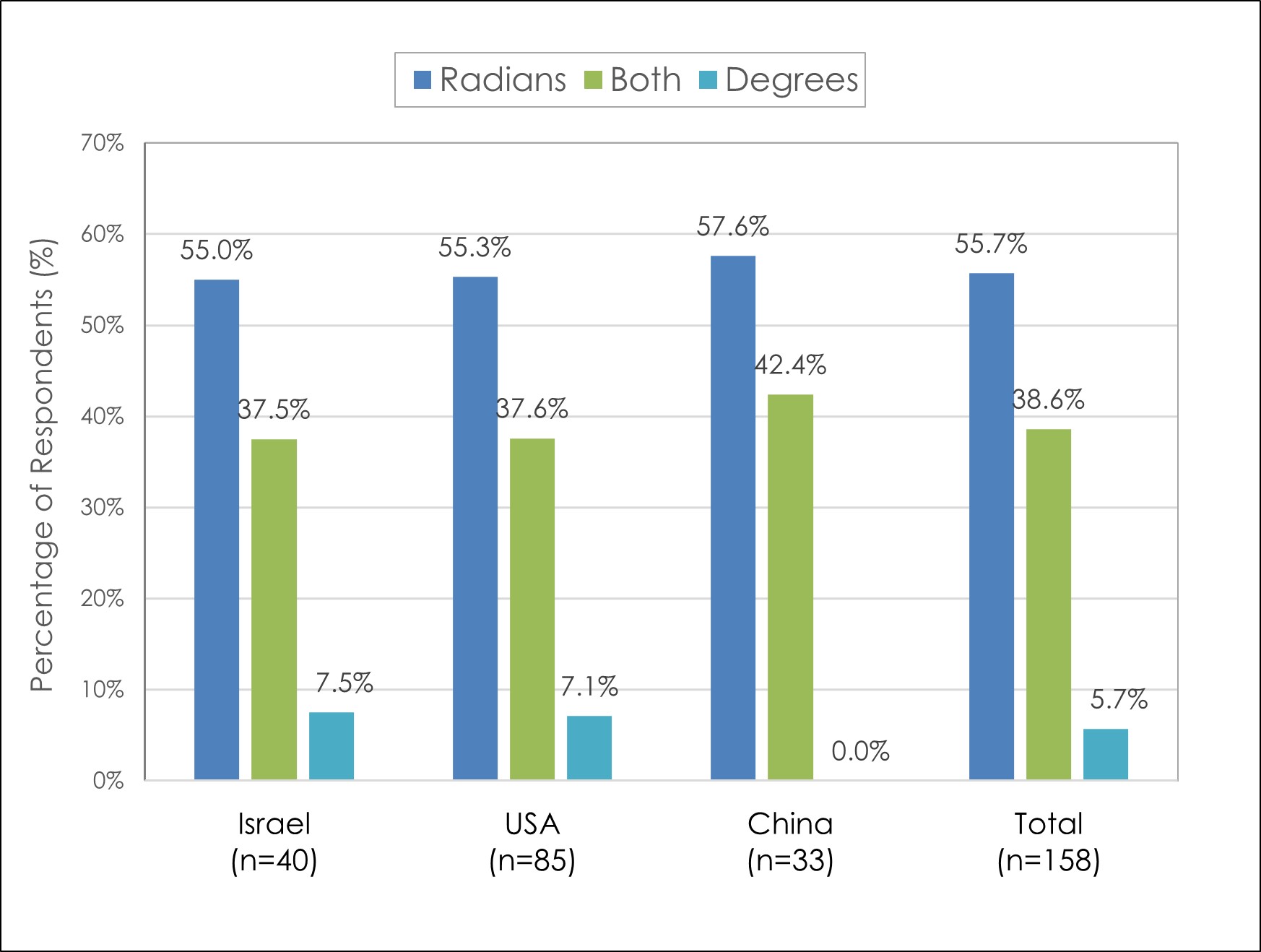}
\caption{\label{fig:matched_Q4} Distribution of responses for Q4 (harmonic oscillator expression) among first- and second-year physics students, across Israel, the United States, and China. Percentages and absolute counts are provided in Table~\ref{tab:34} in the Appendix.}
\end{figure}

To ensure a fair cross-national comparison, we conducted a secondary analysis limited to first- and second-year undergraduate physics students. This matched sample controlled for potential confounding factors related to academic background and curriculum structure. Notably, no students from India met the inclusion criteria, resulting in an analysis that included only students from Israel, the United States, and China. The results, presented in Figures~\ref{fig:matched_Q3} and~\ref{fig:matched_Q4}, reveal strikingly similar response patterns across countries. For Q3, the proportion of students correctly identifying radians as the valid unit for differentiation remained low and exhibited minimal variation between countries (\(\chi^2 (4) = 5.776, p = .216\)). Likewise, for Q4, which assessed students' recognition of radians in the context of harmonic motion, the distribution of responses was nearly identical across the three countries (\(\chi^2 (4) = 5.57, p = .632\)). These findings indicate that understanding the role of radians in differentiation is a widespread cognitive challenge among undergraduate physics students.

\subsection{Analysis of Respondent Types}

We classified the responses by the following criteria:
\begin{enumerate}
    \item \textbf{Consistent Correct:} Respondents who answered both Q3 and Q4 correctly.
    \item \textbf{Q3 Correct Only:} Respondents who answered Q3 correctly but Q4 incorrectly.
    \item \textbf{Q4 Correct Only:} Respondents who answered Q4 correctly but Q3 incorrectly.
    \item \textbf{Consistent Incorrect:} Respondents who answered both Q3 and Q4 incorrectly.
\end{enumerate}

In examining the respondent types by country, a significant association was found \((\chi^2 (9)=17.63, p=.040)\). Results demonstrated that while in Israel and China the consistent correct responses were relatively high (26.1\% in Israel and 21.0\% in China), lower consistent correct responses were found in the USA (15.2\%) and India (15.0\%).

Overall, the highest proportion of respondent types were in the "Q3 incorrect but Q4 correct" category (38.6\%) and the "Q3 and Q4 incorrect" category (35.1\%). About 20\% of the total sample demonstrated consistent correct responses, while only 6\% fell into the "Q3 correct but Q4 incorrect" category.

\begin{figure}[ht]
\centering
\includegraphics[width=0.5\textwidth]{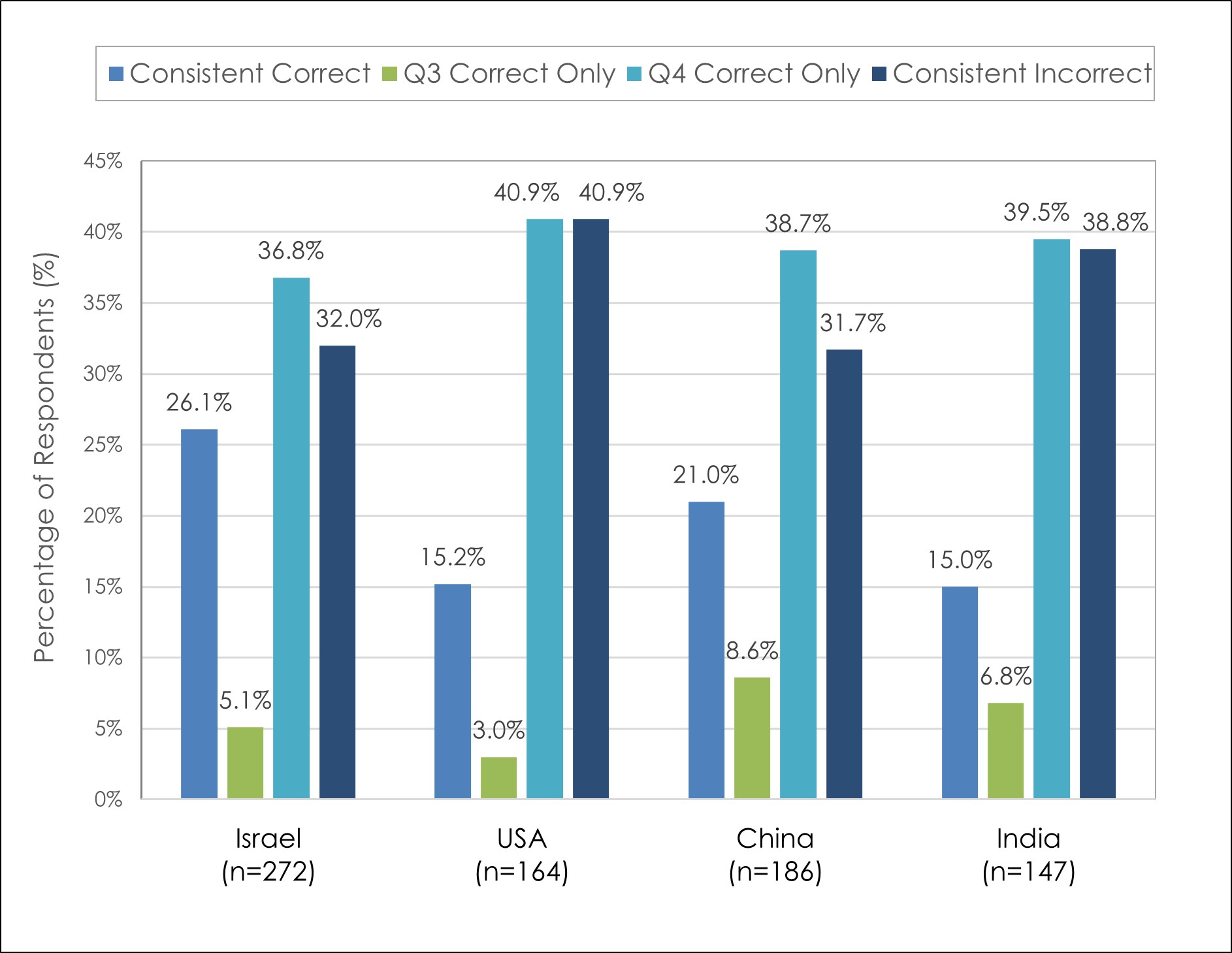}
\caption{\label{fig3:respondent_types} 
Distribution of respondent types across countries. The figure categorizes participants based on their accuracy in answering Q3 (trigonometric derivatives) and Q4 (harmonic oscillator expression). The "Consistent Correct" group is largest in Israel and China, whereas the "Consistent Incorrect" and "Q3 incorrect but Q4 correct" groups dominate in all countries. A detailed breakdown of percentages and absolute counts is provided in Table~\ref{tab:5} in the Appendix.}
\end{figure}

These trends are further illustrated in Figure~\ref{fig3:respondent_types}, which visually represents the distribution of the four respondent types across all countries.

\subsection{Common Challenges in Understanding About Radians in Harmonic Motion}

To analyze the open-ended responses provided for Q5, the answers were categorized into seven distinct types based on their content and reasoning. The categorization process was conducted systematically, as follows:

\begin{enumerate}
   \item \textbf{Derivative Distinction}
\begin{itemize}
    \item \textbf{Description:} Correctly explains that derivatives differ between degrees and radians or explicitly mentions the need for a scaling factor when converting between the two (e.g., recognizing that degrees require adjusting the rate of the function).
    \item \textbf{Significance:} This represents a correct and sophisticated reasoning, indicating an understanding of solving differential equations and the mathematical differences in derivatives.
    \item \textbf{Example:} “Since the derivative of cos is sin only in radians, then the solution must be in radians to satisfy the wave equation.”
\end{itemize}
    \item \textbf{Dimensional Analysis}
\begin{itemize}
    \item \textbf{Description:} Claims that radians must be used for physical or mathematical reasons, often referring to the necessity of specific units. This category includes justifications based on the units of omega and phi.
    \item \textbf{Significance:} This is a correct justification, though it does not necessarily indicate a full understanding of derivatives.
    \item \textbf{Example:} “Omega is angular frequency and is measured in radians/seconds, so I would use radians here.”
\end{itemize}
    \item \textbf{Units Equivalence}
\begin{itemize}
    \item \textbf{Description:} Claims that both degrees and radians can be used interchangeably, emphasizing conversion between them without adjusting constants like omega or phi. This type includes transitions that do not account for the need to adjust the constants themselves when changing the units.
    \item \textbf{Significance:} This is an incorrect justification that reveals a misunderstanding of the distinction between radians and degrees.
    \item \textbf{Example:} “Both are correct, as they are essentially a measurement of the same thing.”
\end{itemize}
    \item \textbf{Experience/Authority-Based}
\begin{itemize}
    \item \textbf{Description: }Relies on personal experience, memory, or authority, such as referencing past instruction or standard practices. Responses often include statements like, “That’s how we’ve always done it,” or mention what was taught by a teacher.
    \item \textbf{Significance:} This reasoning is often unsupported by a clear understanding of the underlying mathematical principles.
    \item \textbf{Example: }“I remember when I learned about circular motion (which was then applied to simple harmonic motion), I was told that units of omega are radians per second.”
\end{itemize}
    \item \textbf{Unsupported Statement}
\begin{itemize}
    \item \textbf{Description: }Provides a statement without further explanation or justification.
    \item \textbf{Significance: }While sometimes technically correct, these responses lack depth and clarity.
    \item \textbf{Example:} “Radians are the natural units.”
\end{itemize}
    \item \textbf{Explicit Uncertainty}
\begin{itemize}
    \item \textbf{Description:} Admits a lack of knowledge or states that the answer was guessed.
    \item Significance: Reflects uncertainty or a lack of engagement with the question.
    \item \textbf{Example:} “I guessed.”
\end{itemize}
    \item \textbf{No Response}
\begin{itemize}
    \item \textbf{Description:} Provides no answer or an irrelevant response, including challenges in understanding such as “only radians are unitless.”
    \item \textbf{Significance:} Indicates a failure to engage with the question or a fundamental misunderstanding.
\end{itemize}
\end{enumerate}
The coding process was conducted blindly by the first author and two additional independent reviewers. The final coding for each response was determined by the majority agreement among the three coders, except in 11 cases out of 770 (1.43\%), where the author opted for a different classification based on contextual analysis. Inter-rater reliability was assessed using Cohen's Kappa coefficient, which yielded moderate (0.56) to high values of (0.76) between the coders, indicating robust consistency among the reviewers. 
The Cohen’s Kappa values (0.56 to 0.76) reflect the initial agreement levels among coders before applying majority-rule decisions. This indicates that while agreement was generally high, there was still some variation in initial classifications that was resolved through consensus.

Table \ref{tab:6} presents the distribution of common challenges in understanding about radians in harmonic motion. The most common explanation, given by 23.9\% of participants (184), was the Units Equivalence category, which claims that both methods can be used and emphasizes the possibility of converting between them. 
Dimensional Analysis, which asserts that radians must be used for physical or mathematical reasons and refers to the nature and necessity of units, was cited by 17\% of participants (131). Unsupported Statements, consisting of explanations without any justification, accounted for 13.5\% (104). Experience/Authority-Based reasoning, where participants relied on personal experience, teacher authority, or tradition, was given by 10.4\% (80).
Additionally, 10.3\% (79) of participants expressed Explicit Uncertainty, admitting a lack of knowledge or guessing. Finally, the least common theme was Derivative Distinction, mentioned by 6.4\% (49), which reflected an understanding that derivatives differ between degrees and radians.

\begin{table}[ht]
\caption{\label{tab:6}
Categories of Student Reasoning About Radians in Harmonic Motion}
\begin{ruledtabular}
\begin{tabular}{lcc}
Explanation & N & \% \\
\hline
Derivative Distinction\(^*\) & 49 & 6.4 \\
Dimensional Analysis\(^*\) & 131 & 17 \\ 
Units Equivalence & 184 & 23.9 \\
Experience/Authority Based & 80 & 10.4 \\
Unsupported Statement & 104 & 13.5 \\
Explicit Uncertainty & 79 & 10.3 \\
No Response & 142 & 18.5 \\ 
\end{tabular}
\end{ruledtabular}
\begin{flushleft}
\textit{Note:} Categories marked with an asterisk (\(^*\)) indicate correct reasoning.
\end{flushleft}
\end{table}

The reasoning provided by students was divided into three main categories:
\begin{enumerate}
    \item \textbf{Meaningful Reasoning:} These responses demonstrate a correct or near-correct understanding of the physical or mathematical principles involved, even if the explanation is not entirely precise. They reflect a deeper engagement with the problem:
\begin{itemize}
    \item \textbf{Category 1 - Derivative Distinction:} Arguments recognizing the difference between derivatives in degrees and radians and how this distinction impacts the physical system.
    \item \textbf{Category 2 - Dimensional Analysis:} Arguments based on the nature of radians as dimensionless units and their necessity in physical systems.

\end{itemize}
    \item \textbf{Misguided Reasoning:} These responses attempt to explain the choice but rely on incorrect assumptions or misunderstandings of the principles. While incorrect, they show an effort to think critically about the problem:
\begin{itemize}
    \item \textbf{Category 3 - Units Equivalence:} Claims that degrees and radians are interchangeable, indicating a conceptual error but a logical attempt to rationalize the response.
\end{itemize}
    \item \textbf{Meaningless Reasoning:} These responses fail to provide meaningful insights into the student's understanding, often due to a lack of knowledge, unsupported statements, or irrelevant answers:
\begin{itemize}
    \item \textbf{Category 4 - Experience/Authority-Based:} Responses relying on past experiences or authority ("that's how we were taught") without further explanation.
    \item \textbf{Category 5 - Unsupported Statement:} General statements with no basis or explanation.
    \item \textbf{Category 6 - Explicit Uncertainty:} Admissions of uncertainty or guesses.
    \item \textbf{Category 7 - No Response:} Blank or irrelevant responses.
\end{itemize}
\end{enumerate}

This categorization is valuable for both qualitative and quantitative analysis. Meaningful reasoning provides insight into genuine understanding, misguided reasoning highlights conceptual difficulties, and meaningless reasoning identifies gaps in knowledge or engagement. By analyzing these categories, we can better understand student reasoning patterns and target specific areas for educational improvement.

To further examine the distribution of reasoning types across different educational systems, we analyzed the classification of reasoning by country. A significant association was found \((\chi^2 (6)=19.25, p=.004)\), indicating meaningful differences between nations.

\begin{figure}[ht]
\centering
\includegraphics[width=0.5\textwidth]{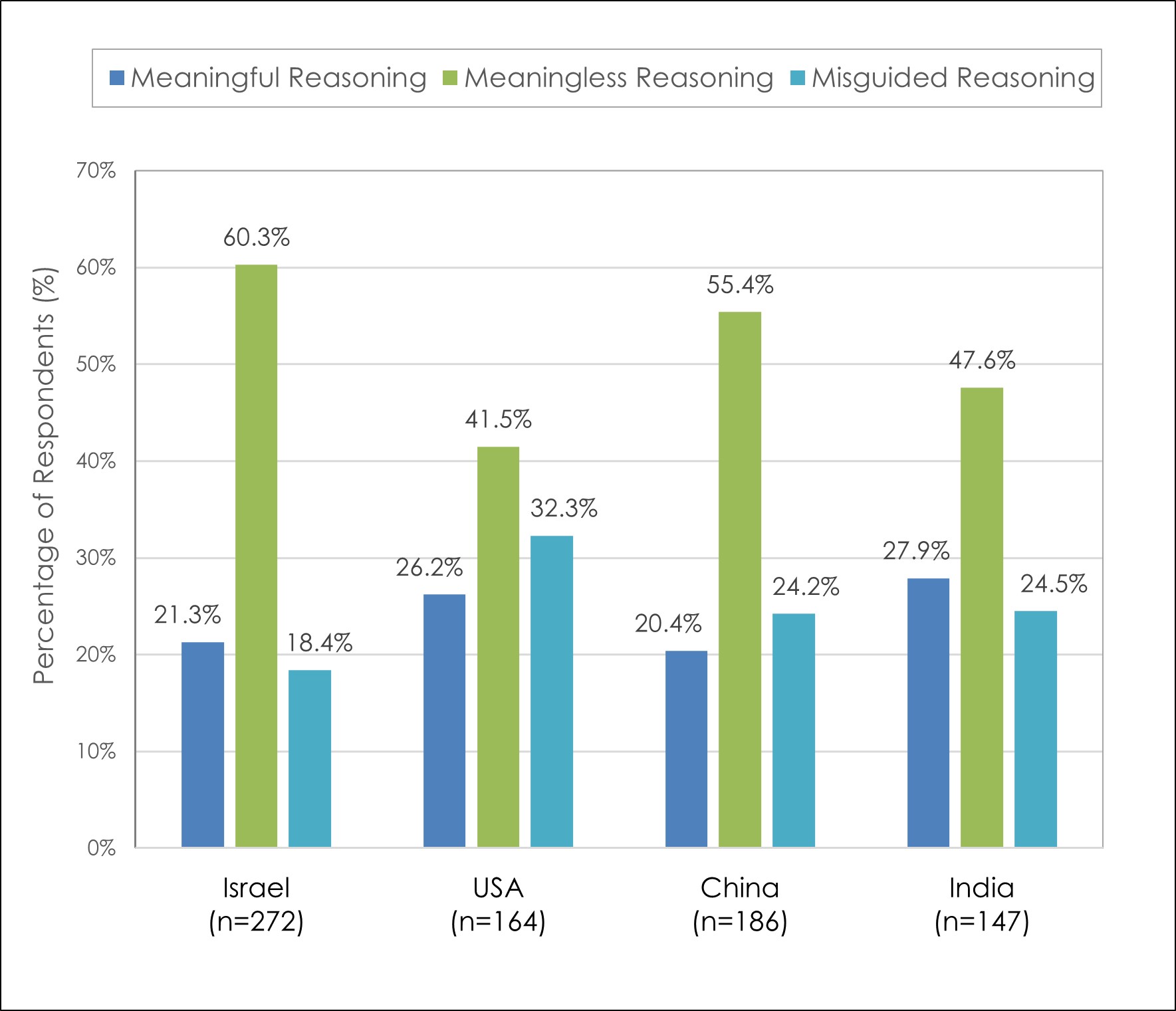}
\caption{\label{fig4:reasoning_distribution} 
Distribution of reasoning types across countries. The figure categorizes participants based on their reasoning classification: Meaningful, Misguided, and Meaningless reasoning .A detailed breakdown of percentages and absolute counts is provided in Table~\ref{tab:7} in the Appendix.}
\end{figure}

As shown in Figure~\ref{fig4:reasoning_distribution}, students in the United States (26.2\%) and India (27.9\%) exhibited the highest rates of meaningful reasoning, whereas lower rates were observed in Israel (21.3\%) and China (20.4\%). Conversely, the proportion of students classified under meaningless reasoning was highest in Israel (60.3\%) and China (55.4\%), while the lowest rates were observed in the USA (41.5\%) and India (47.6\%). Misguided reasoning responses were relatively balanced across countries, ranging from 18.4\% in Israel to 32.3\% in the USA.

\subsection{Confidence Level}

We conducted a one-way ANOVA to compare the confidence levels (Q6) among students based on their responses to the Correct Case Code for the Harmonic Oscillator Expression (Q4). The results indicated a significant difference between the response groups\( (F(2,766) = 12.35, p < .001, \eta^2 = 0.03)\), suggesting that students' confidence levels vary depending on their conceptual understanding.

Post hoc Tukey HSD analysis revealed that the lowest confidence level was among students who incorrectly responded "Angle in degrees", compared to those who selected either "Both (Incorrect)" or "Radians (Correct)" \((p < .001)\). Importantly, there was no significant difference in confidence levels between students who answered correctly ("Radians (Correct)") and those who incorrectly believed both units were valid ("Both (Incorrect)", \(p = .95\)).
These findings suggest that students who misunderstand the fundamental distinction between radians and degrees in the context of harmonic motion may still exhibit high confidence in their incorrect responses, reinforcing the hypothesis of a global "blind spot" in trigonometric differentiation.

Table~\ref{tab:8} presents the mean confidence levels for each response type, on a 5-point Likert scale, with 5 being most confident. Students who selected "Angle in degrees" exhibited the lowest confidence, while those who incorrectly chose "Both (Incorrect)" maintained confidence levels comparable to correct responses.

\begin{table}[ht]
\caption{\label{tab:8}
Confidence level by response}
\begin{ruledtabular}
\begin{tabular}{lcccc}
 Correct Case Code for  \\
       Harmonic Oscillator \\
     Expression  & M & SD \\
\hline
Radians & 3.78 & 1.09 \\
Both  & 3.76 & 1.08 \\
Degrees  & 2.96 & 1.34 \\ 
\hline
F (2,766)  & \multicolumn{2}{c}{12.35} \\ 
P  & \multicolumn{2}{c}{\(<.001\)} \\
Effect size (Eta square) & \multicolumn{2}{c}{.03} \\ 
\end{tabular}
\end{ruledtabular}
\end{table}

In comparing the confidence level by respondent types, no significant association was found \((F(2,765)=1.80, p=.1451, \eta^2=0.007)\), which means that all three respondent types demonstrated a similar moderate level of confidence. The descriptive statistics for confidence levels across respondent types are summarized in Table \ref{tab:9}, showing no substantial differences across groups.

\begin{table}[ht]
\caption{\label{tab:9}
Confidence level by respondent types}
\begin{ruledtabular}
\begin{tabular}{lcccc}
& M & SD \\
\hline
Consistent Correct & 3.89 & 1.04 \\
Q3 Correct Only & 3.69 & 1.24 \\
Q4 Correct Only & 3.73 & 1.11 \\
Consistent Incorrect & 3.63 & 1.15 \\
\hline
F (2,766) & \multicolumn{2}{c}{1.80} \\ 
P & \multicolumn{2}{c}{.145} \\
Effect size (Eta square) & \multicolumn{2}{c}{.007} \\ 
\end{tabular}
\end{ruledtabular}
\end{table}

An analysis of confidence levels across reasoning categories revealed a significant effect \((F(2,766) = 21.70, p < .001, \eta^2 = 0.054)\). Post hoc comparisons using the Sidak correction indicated that students who provided meaningless reasoning reported the lowest confidence \((M = 3.48, SD = 1.20)\) compared to those who provided meaningful reasoning \((M = 4.04, SD = 0.94, p < .001)\) or misguided reasoning \((M = 3.94, SD = 0.96, p < .001)\). However, no significant difference in confidence was found between students who provided meaningful versus misguided reasoning \((p = .95)\), suggesting that even incorrect but structured reasoning can sustain high confidence levels. These findings are summarized in Table \ref{tab:10}, which presents the descriptive statistics for confidence levels across reasoning categories.

\begin{table}[ht]
\caption{\label{tab:10}
Confidence level by reasoning}
\begin{ruledtabular}
\begin{tabular}{lcc}
 & M & SD \\ 
\hline
Meaningful Reasoning & 4.04 & 0.938 \\
Misguided Reasoning & 3.94 & 0.965 \\
Meaningless Reasoning & 3.48 & 1.197 \\ 
\hline
F (2,766) & \multicolumn{2}{c}{21.70} \\
p & \multicolumn{2}{c}{\(<.001\)} \\
Effect size (Eta square) & \multicolumn{2}{c}{.054} \\
\end{tabular}
\end{ruledtabular}
\end{table}

\subsection{The Correlation between Calculus Grades and Students' Answers to Derivatives}

We examined whether students' self-reported scores in Calculus are correlated with the correctness of their answers in the Trig Derivatives question (Q3).  

Results showed that no significant association was found between the grouped calculus score and the correctness of the answers in the USA \((\chi^2 (2) = 1.147, p = .564)\), China \((\chi^2 (2) = 0.915, p = .633)\), or India \((\chi^2 (2) = 1.221, p = .543)\). However, a significant association was found in Israel \((\chi^2 (2) = 7.563, p = .023)\), suggesting that higher calculus scores were associated with improved performance in this country only.  

These findings are summarized in Table \ref{Calculus Grades}, which presents the distribution of correct and incorrect answers by grouped calculus score categories across all four countries.

\begin{table*}
\caption{\label{Calculus Grades}
Correctness of the Students' Responses for Trig Derivatives (Q3) by Grouped Calculus Score in All Countries}
\begin{ruledtabular}
\begin{tabular}{lcccccccc}
 & \multicolumn{2}{c}{Israel} & \multicolumn{2}{c}{USA} & \multicolumn{2}{c}{China} & \multicolumn{2}{c}{India} \\
 Calculus Score & Incorrect & Correct & Incorrect & Correct & Incorrect & Correct & Incorrect & Correct \\
\hline
Low (0-79)  & 78.9\% (30) & 21.1\% (8) & 84.6\% (11) & 15.4\% (2) & 74.1\% (43) & 25.9\% (15) & 86.7\% (13) & 13.3\% (2) \\
Medium (80-94)  & 73.9\% (85) & 26.1\% (30) & 85.7\% (60) & 14.3\% (10) & 67.6\% (71) & 32.4\% (34) & 78.6\% (66) & 21.4\% (18) \\
High (95-100)  & 59.8\% (70) & 40.2\% (47) & 78.6\% (44) & 21.4\% (12) & 73.9\% (17) & 26.1\% (6) & 70.6\% (12) & 29.4\% (5) \\

\hline
$\chi^2$ (2)  & \multicolumn{2}{c}{7.563} & \multicolumn{2}{c}{1.147} & \multicolumn{2}{c}{0.915} & \multicolumn{2}{c}{1.221} \\
P  & \multicolumn{2}{c}{.023} & \multicolumn{2}{c}{.564} & \multicolumn{2}{c}{.633} & \multicolumn{2}{c}{.543} \\
\end{tabular}
\end{ruledtabular}
\end{table*}

\subsection{The Correlation between Mechanics Grades and Students' Answers to Harmonic Oscillator Questions}

We examined whether students' self-reported scores in Mechanics are correlated with the correctness of their answers in the Harmonic Oscillator Expression question (Q4).  

Results showed that no significant association was found between the grouped mechanics score and correctness in Israel \((\chi^2 (2) = 2.898, p = .235)\), the USA \((\chi^2 (2) = 1.949, p = .377)\), or China \((\chi^2 (2) = 2.777, p = .249)\). However, a significant association was found in India \((\chi^2 (2) = 6.087, p = .048)\), suggesting that higher mechanics scores were associated with improved performance in this country only.  

These findings are summarized in Table \ref{tab:12}, which presents the distribution of correct and incorrect answers by grouped mechanics score categories across all four countries.

\begin{table*}
\caption{\label{tab:12}
Correctness of the Students' Responses for the Harmonic Oscillator Expression (Q4) by Grouped Mechanics Score in All Countries}
\begin{ruledtabular}
\begin{tabular}{lcccccccc}
 & \multicolumn{2}{c}{Israel} & \multicolumn{2}{c}{USA} & \multicolumn{2}{c}{China} & \multicolumn{2}{c}{India} \\
 Mechanics Score & Incorrect & Correct & Incorrect & Correct & Incorrect & Correct & Incorrect & Correct \\
\hline
Low (0-79)  & 39.4\% (28) & 60.6\% (43) & 51.9\% (14) & 48.1\% (13) & 51.5\% (17) & 48.5\% (16) & 66.7\% (14) & 33.3\% (7) \\
Medium (80-94) & 31.7\% (39) & 68.3\% (84) & 44.2\% (34) & 55.8\% (43) & 36.4\% (40) & 63.6\% (70) & 40.4\% (36) & 59.6\% (53) \\
High (95-100) & 43.2\% (32) & 56.8\% (42) & 35.0\% (14) & 65.0\% (26) & 33.3\% (7) & 66.7\% (14) & 27.3\% (3) & 72.7\% (8) \\
\hline
\(\chi^2\) (2) &\multicolumn{2}{c}{2.898} & \multicolumn{2}{c}{1.949} &\multicolumn{2}{c}{2.777}  & \multicolumn{2}{c}{6.087} \\
P & \multicolumn{2}{c}{.235} & \multicolumn{2}{c}{.377} & \multicolumn{2}{c}{.249} & \multicolumn{2}{c}{.048}  \\
\end{tabular}
\end{ruledtabular}
\end{table*}

\subsection{The Relationship Between Academic Departments and Students' Answers}

We examined whether the academic department (field of study) is associated with the students' answers to Q3 (correctness of identifying derivatives) and Q4 (correctness of identifying the correct condition for a harmonic oscillator).

Results showed a significant association between the academic department and the correctness of the answers in the Trig Derivatives question \((\chi^2 (4)=22.26, p<.001)\) with the highest rate of correct responses among Engineering (39.30\%), and the lowest rate of correct responses among Chemistry, Life Sciences, or Neuroscience (17.10\%).

Results showed a significant association between the academic department and the correctness of the answers in the Harmonic Oscillator Expression question \((\chi^2 (4)=13.72, p<.001)\) with the highest rate of correct responses among Engineering (69.60\%) and Math and Computer Science (66.10\%), while the lowest rate of correct responses among Physics and any other field (54.40\%) and Chemistry, Life Sciences, or Neuroscience (54.10\%). As shown in Table~\ref{tab:13}, the rates of correct answers differ significantly across academic departments.

\begin{table*}
\caption{\label{tab:13}
Correctness of the Students' Responses by Academic Department}
\begin{ruledtabular}
\begin{tabular}{lcccc}
Academic Department & \multicolumn{2}{c}{Trig Derivatives Question} & \multicolumn{2}{c}{Harmonic Oscillator Expression Question} \\ 
 & Incorrect & Correct & Incorrect & Correct \\ 
 \hline
Physics& 77.2\% (298) & 22.8\% (88) & 45.6\% (176) & 54.4\% (210) \\ 
Engineering & 60.7\% (102) & 39.3\% (66) & 30.4\% (51) & 69.6\% (117) \\ 
Math or Computer Science & 71.0\% (44) & 29.0\% (18) & 33.9\% (21) & 66.1\% (41) \\ 
Chemistry, Life Sciences, or Neuroscience & 82.9\% (92) & 17.1\% (19) & 45.9\% (51) & 54.1\% (60) \\ 
Other & 75.6\% (31) & 24.4\% (10) & 39.0\% (16) & 61.0\% (25) \\ \hline
\(\chi^2\) (4) & \multicolumn{2}{c}{22.26} & \multicolumn{2}{c}{13.72} \\ 
P & \multicolumn{2}{c}{\(<.001\)} & \multicolumn{2}{c}{\(<.001\)} \\ 
\end{tabular}
\end{ruledtabular}
\end{table*}

\subsection{Multivariate logistic regression predicting correct responses}

To assess the relationship between multiple independent variables and response accuracy, we conducted multivariate binary logistic regression analyses. Table~\ref{tab:14} presents the results.

\subsubsection{Predicting Correct Responses for Trigonometric Derivatives (Q3)}  
The analysis revealed that engineering students had significantly higher odds of answering correctly compared to chemistry students (\(\text{OR} = 3.253\), 95\% CI: 1.854–5.709, \(p < 0.001\)). Mathematics and Computer Science students also exhibited significantly higher odds of success (\(\text{OR} = 2.579\), 95\% CI: 1.166–5.707, \(p = 0.019\)). Physics students were more likely to succeed (\(\text{OR} = 2.011\), 95\% CI: 1.143–3.539, \(p = 0.015\)).  

Students from the USA had significantly lower odds of answering correctly compared to students from Israel (\(\text{OR} = 0.387\), 95\% CI: 0.215–0.697, \(p = 0.002\)). No significant differences were observed for students from China or India compared to Israel (\(p > 0.05\)).  

No significant effects were found for GPA (\(p = 0.225\)), mechanics scores (\(p = 0.317\)), or calculus scores (\(p = 0.078\)). Academic year (first three years vs. later years) was also not a significant predictor (\(\text{OR} = 1.676\), 95\% CI: 0.826–3.403, \(p = 0.153\)).  

The model correctly classified 73.7\% of cases.

\subsubsection{Predicting Correct Responses for the Harmonic Oscillator Expression (Q4)}  
The analysis showed that students with a GPA of 95+ had significantly higher odds of answering correctly compared to those with lower GPAs (\(\text{OR} = 2.361\), 95\% CI: 1.259–4.426, \(p = 0.007\)).  

Engineering students had significantly higher odds of answering correctly compared to chemistry students (\(\text{OR} = 1.884\), 95\% CI: 1.129–3.142, \(p = 0.015\)). Mathematics and Computer Science students showed no significant difference compared to chemistry students (\(p = 0.109\)).  

Response confidence level was a significant predictor of accuracy. A one-unit increase in confidence was associated with higher odds of responding correctly (\(\text{OR} = 1.229\), 95\% CI: 1.051–1.438, \(p = 0.010\)).  

No significant effects were found for country of origin (\(p > 0.05\)), mechanics scores (\(p = 0.439\)), calculus scores (\(p = 0.682\)), or physics vs. chemistry (\(p = 0.808\)). Academic year was not a significant predictor (\(\text{OR} = 1.280\), 95\% CI: 0.703–2.330, \(p = 0.420\)).  

The model correctly classified 63.0\% of cases.

\begin{table*}
\caption{\label{tab:14} Multivariate logistic regression predicting correct responses}
\begin{ruledtabular}
\begin{tabular}{lcccccccc}
 & \multicolumn{4}{c}{Trig Derivatives (Q3)} & \multicolumn{4}{c}{Harmonic Oscillator Expression (Q4)} \\ 
 & OR & \multicolumn{2}{c}{95\% CI} & p & OR & \multicolumn{2}{c}{95\% CI} & p \\
  & & Lower & Upper & & & Lower & Upper & \\
\hline
US vs. Israel & 0.387 & 0.215 & 0.697 & 0.002 & 0.776 & 0.473 & 1.272 & 0.314 \\
China vs. Israel & 0.983 & 0.599 & 1.612 & 0.945 & 0.745 & 0.466 & 1.193 & 0.220 \\
India vs. Israel & 0.976 & 0.431 & 2.212 & 0.954 & 0.950 & 0.460 & 1.965 & 0.891 \\
GPA 95+ vs. Below 95 & 1.487 & 0.784 & 2.821 & 0.225 & 2.361 & 1.259 & 4.426 & 0.007 \\
Mechanics 95+ vs. Below 95 & 1.301 & 0.777 & 2.179 & 0.317 & 0.830 & 0.519 & 1.330 & 0.439 \\
Calculus 95+ vs. Below 95 & 1.473 & 0.958 & 2.263 & 0.078 & 1.087 & 0.730 & 1.618 & 0.682 \\
Physics vs. Chemistry & 2.011 & 1.143 & 3.539 & 0.015 & 0.941 & 0.576 & 1.537 & 0.808 \\
Engineering vs. Chemistry & 3.253 & 1.854 & 5.709 & \(<0.001\) & 1.884 & 1.129 & 3.142 & 0.015 \\
Math \& CS vs. Chemistry & 2.579 & 1.166 & 5.707 & 0.019 & 1.805 & 0.877 & 3.714 & 0.109 \\
Bachelor's Years 1-3 vs. 4+ & 1.676 & 0.826 & 3.403 & 0.153 & 1.280 & 0.703 & 2.330 & 0.420 \\
Response Confidence Level & - & - & - & - & 1.229 & 1.051 & 1.438 & 0.010 \\
\hline
\% correct classification & \multicolumn{4}{c}{73.7\%} & \multicolumn{4}{c}{63.0\%} \\ 
\end{tabular}
\end{ruledtabular}
\end{table*}

\subsection{Multivariate linear regression predicting confidence level}

Table~\ref{tab:16} presents the results of the multivariate linear regression predicting confidence levels. Several significant associations were identified.

Compared to Israel, students from China (\(\beta = 0.597, p < .001\)) exhibited significantly higher confidence levels, and a similar trend was observed for students from India (\(\beta = 0.350, p = 0.043\)). No significant difference was found between students from the United States and Israel.

Field of study also played a role: students majoring in Physics (\(\beta = 0.509, p < .001\)) and Engineering (\(\beta = 0.397, p = 0.001\)) demonstrated higher confidence compared to those in Chemistry, while no significant effect was found for students majoring in Mathematics and Computer Science (\(\beta = 0.149, p = 0.384\)).

A key finding was the strong negative association between explicit uncertainty and confidence level (\(\beta = -1.114, p < .001\)), indicating that students who explicitly expressed uncertainty were substantially less confident in their responses. Additionally, students who incorrectly selected "both are correct" as a valid unit for the Harmonic Oscillator expression exhibited significantly higher confidence compared to those who correctly identified "radians" (\(\beta = 0.623, p = 0.005\)). Similarly, students who correctly selected "radians" showed significantly higher confidence compared to those who selected "degrees" (\(\beta = 0.707, p = 0.002\)).

No significant effects were found for GPA (\(\beta = 0.021, p = 0.879\)), mechanics scores (\(\beta = 0.119, p = 0.290\)), calculus scores (\(\beta = 0.048, p = 0.613\)), or academic year (\(\beta = -0.130, p = 0.366\)), suggesting that these factors were not major predictors of confidence level.

The overall model explained 25.8\% of the variance in confidence levels (\( R^2 = 0.258 \)), with a significant overall model fit (\( F(13, 635) = 16.989, p < .001 \)).

\begin{table}[ht]
\caption{\label{tab:16} Multivariate linear regression predicting confidence level}
\begin{ruledtabular}
\begin{tabular}{lcc}
Variable & Beta & p \\
\hline
US vs. Israel & -0.146 & 0.215 \\
China vs. Israel & 0.597 & \(<.001\) \\
India vs. Israel & 0.350 & 0.043 \\
GPA 95+ vs. lower & 0.021 & 0.879 \\
Mechanics 95+ vs. lower & 0.119 & 0.290 \\
Calculus 95+ vs. lower & 0.048 & 0.613 \\
Physics vs. Chemistry & 0.509 & \(<.001\) \\
Engineering vs. Chemistry & 0.397 & 0.001 \\
Math \& CS vs. Chemistry & 0.149 & 0.384 \\
Explicit Uncertainty vs. other explanations & -1.114 & \(<.001\) \\
Q4 Radians vs. Degrees & 0.707 & 0.002 \\
Q4 Both Correct vs. Degrees & 0.623 & 0.005 \\
Undergrad (1-3 BA only) vs. Advanced Studies & -0.130 & 0.366 \\
\end{tabular}
\end{ruledtabular}
\end{table}

\section{Disscussion}

This study investigated a global "blind spot" in understanding trigonometric derivatives and their implications for physics education. The analysis focused on students' understanding of the distinction between radians and degrees, their ability to apply this knowledge to physical systems such as harmonic oscillators, the associations between confidence and performance, and the prevalence of these issues across different educational systems and countries. Below, we address each research question and discuss the broader implications.

\subsection{Do students correctly identify the derivatives of trigonometric functions in radians versus degrees?}

The results demonstrate a significant gap in students' understanding of the conditions under which trigonometric derivatives are valid in an abstract / mathematical context. Only 26.3\% of participants correctly identified that the known derivative of a trigonometric function is valid exclusively in radians. The majority (70.7\%) incorrectly assumed that both radians and degrees are equally valid, indicating a widespread misunderstanding of the mathematical distinction. This error is critical, as it reflects a failure to recognize the scaling factor required when differentiating in degrees.

\subsection{Do students understand the difference between radians and degrees in physical systems such as a harmonic oscillator?}

Approximately 59.0\% of participants correctly identified that the standard harmonic oscillator equation assumes angle measurements in radians. However, a substantial proportion (34.7\%) believed that both radians and degrees are valid. This misunderstanding is particularly problematic in physics, where incorrect assumptions about units can lead to significant errors in modeling and interpretation.

It is important to evaluate the results of these two research questions in the context of previous studies conducted within a similar framework, where the ability of students to comprehend and transfer mathematical and physical concepts was investigated. These studies reveal common patterns of difficulty, along with unique differences in their outcomes, offering valuable insights into the interplay between mathematics and physics.

Beichner \cite{beichner1994testing} examined students’ understanding of kinematics graphs (position-time and velocity-time) using a dedicated assessment tool (TUG-K). The findings revealed that students struggled to connect slopes with their physical meaning, such as velocity. Physical graphs introduced confusion, as students found it challenging to translate mathematical understanding into physical concepts.

Christensen and Thompson \cite{christensen2012investigating} focused on students' ability to interpret derivatives and slopes in a purely mathematical context, without introducing physical elements. It revealed that many students lacked a deep understanding of the relationship between the derivative and the slope, even in a mathematical context. Difficulties in physics often stem from a lack of foundational understanding of derivatives in mathematics.

Carli et al. \cite{PhysRevPhysEducRes.16.010111} assessed students' ability to solve isomorphic questions in mathematics and physics, showing that students performed better in purely mathematical questions than in physical ones. This was attributed to the cognitive bridge required to transfer knowledge between the two domains. Physical contexts impose an additional cognitive demand that many students struggle to manage effectively.

In contrast to previous studies, our results showed that students were more successful in identifying the necessity of using radians in physical questions than in mathematical ones. This finding highlights that students, within a physical context, demonstrated a higher recognition rate for the dependency on radians.

The uniqueness of our findings lies in what we term the physical anchor. The familiarity of the students with physical systems, such as the harmonic oscillator, allowed them to recognize that commonly used expressions (for example, $\omega$, $x(t)=Acos(\omega t+\phi)$) are valid only when trigonometric functions are expressed in radians. This suggests that in some cases physical context provided a support structure for students to correctly identify the relevant mathematical framework.

However, this improved recognition does not necessarily reflect a deep mathematical understanding. Many students relied on their memory of familiar physical expressions rather than systematically analyzing the difference between radians and degrees.

In light of Gifford and Finkelstein’s \cite{PhysRevPhysEducRes.16.020121,PhysRevPhysEducRes.17.010138}  framework on mathematical sense-making (MSM), our findings may reflect a distinction in how students engage with mathematical concepts in different contexts. Specifically, in the mathematical setting, students must apply derivatives as a tool to make sense of another mathematical object (e.g., differentiating \( \cos(\theta) \)), a process categorized as MSM-M (Mathematical Sense-Making in Mathematics). In contrast, in the physics context, derivatives serve as a tool to interpret a physical system (e.g., the harmonic oscillator), aligning with MSM-P (Mathematical Sense-Making in Physics). This shift in reasoning suggests that students may perceive these as fundamentally different types of problems, activating distinct cognitive resources. 

The observed improvement in recognizing radians in physics may stem from students relying on their familiarity with well-established physical equations, rather than a conceptual understanding of why radians must be used. This suggests that their reasoning in the physics context is more procedural, whereas in the mathematical context, they lack a clear conceptual framework to determine when radians are necessary.

\subsection{Are there significant differences in understanding between countries and educational systems?}

The analysis revealed that the rate of incorrect responses to the harmonic oscillator question was similar across countries, ranging between 38.1\% and 45.6\%, with no statistically significant differences. This consistency suggests a fundamental issue that transcends cultural and educational differences, indicating a global blind spot in understanding this concept.

In contrast, the rate of incorrect responses to the trigonometric derivatives question showed statistically significant differences across countries, with incorrect response rates ranging from 68.7\% to 81.7\%. However, the consistently high rates of incorrect answers across all countries underscore the widespread nature of this misunderstanding, reinforcing the notion of a global blind spot rather than a localized problem.

A similar pattern has been observed in previous cross-national studies. For example, Bao et al. \cite{10.1119/1.2976334,doi:10.1126/science.1167740} found that Chinese students outperformed their American counterparts in physics knowledge due to a more structured and intensive curriculum. However, despite these differences in physics education, there was no significant difference between the two groups in their ability to apply scientific reasoning. A parallel emerges in our study: While educational systems differ significantly across countries and while Chinese students typically perform better in structured physics assessments, we find no significant advantage for them in understanding the distinction between radians and degrees. This suggests that, much like the scientific reasoning in the study by Bao et al., the confusion between radians and degrees is not merely a result of differences in educational background but rather a fundamental cognitive blind spot in mathematical reasoning within physics.

\subsection{ What is the correlation between students' confidence levels and the correctness of their answers?}

One of the most intriguing aspects of our findings is the relationship between students' confidence levels and their conceptual understanding. While traditional assessment methods might categorize incorrect responses uniformly as evidence of misunderstanding, a more nuanced analysis of students' confidence levels and reasoning patterns reveals a complex landscape of conceptual difficulties. This complexity becomes particularly evident when viewed through the theoretical lens of confidence-based assessment.

In analyzing students' responses and confidence levels, it is crucial to distinguish between two fundamentally different types of learning difficulties: knowledge gaps and alternative conceptions. Hasan et al. \cite{Saleem_Hasan_1999} proposed a theoretical framework that differentiates between these through the Certainty of Response Index (CRI), where incorrect answers accompanied by low confidence indicate knowledge gaps, while incorrect answers with high confidence suggest the presence of alternative conceptions - robust cognitive structures that differ from accepted scientific understanding.

Our findings align with this theoretical framework in several significant ways. Students who incorrectly identified trigonometric derivatives as valid in both radians and degrees demonstrated high confidence levels \((M = 3.76, SD = 1.08)\) statistically indistinguishable from those who answered correctly \((M = 3.78, SD = 1.09)\). According to Hasan et al.'s model, this pattern is characteristic of alternative conceptions rather than mere knowledge deficiency. 

Further support for this theoretical framework emerges from the analysis of students' explanations. Those providing meaningless reasoning exhibited the lowest confidence levels (M = 3.48, SD = 1.20), while students offering meaningful explanations, whether correct or incorrect, displayed significantly higher confidence (\(M = 4.04, SD = 0.94\) and \(M = 3.94, SD = 0.96\), respectively). This pattern aligns with the CRI framework's distinction between knowledge gaps (manifested in low confidence) and alternative conceptions (exhibited through high confidence regardless of correctness). 

These findings indicate that students' difficulty in understanding the role of radians in trigonometric derivatives stems from a strong commitment to the equivalence of degrees and radians, rather than a simple knowledge gap. The consistently high confidence levels among students who believe in the mathematical equivalence of radians and degrees, coupled with their ability to provide meaningful (albeit incorrect) justifications, suggest a deeply ingrained alternative conceptual framework. The persistence of this pattern across four distinct educational systems and cultural contexts further underscores the fundamental nature of this alternative conception in students' mathematical reasoning.

This confidence-maintenance effect is not limited to trigonometric derivatives. A similar pattern has been observed in mechanics education, where alternative conceptions persist despite students' strong confidence in their reasoning. Research on students' understanding of force and motion has shown that misconceptions in mechanics are often accompanied by high confidence, reinforcing the idea that such errors are not merely the result of missing knowledge but rather stem from deeply ingrained alternative conceptions \cite{potgieter2010confidence}.

\subsection{Additional Considerations Regarding Confidence Levels}

The overall confidence level reported was moderate (\(M = 3.72, SD = 1.11\)). One possible explanation for this finding is that participants were required to justify their answers before rating their confidence. Previous research in mathematical problem solving suggests that requiring people to provide explanations before making confidence judgments tends to lower reported confidence levels, although the general tendency towards overestimation persists \cite{wright2024justifying}. This interpretation is further supported by our linear regression analysis, which identifies explicit uncertainty statements as the most significant predictor of reduced confidence. Specifically, participants who explicitly expressed uncertainty exhibited a substantial decrease in confidence (\(\beta = -1.067, p < .001\)), with this predictor demonstrating both the strongest statistical significance and the highest absolute \(\beta\) value in the model. This suggests that the requirement for verbal justification played a central role in moderating confidence levels across the sample.

Another noteworthy trend emerges when comparing confidence levels across countries. Students from China and India exhibited higher confidence levels than their counterparts in Israel and the United States. This pattern aligns with large-scale studies on mathematical problem-solving\cite{stankov2014overconfidence}, which have found that Chinese and Indian students tend to report higher confidence compared to American students. In China, this confidence is also associated with superior performance due to the highly structured and rigorous secondary education system. However, in India, confidence levels were higher without a corresponding advantage in performance. As Bao’s extensive research suggests \cite{10.1119/1.2976334,doi:10.1126/science.1167740}, Chinese students may perceive the problem as relatively simple due to their extensive training in mathematics and physics. However, because this particular question involves a deeply rooted alternative conception rather than a mere procedural skill, confidence levels do not correlate with correctness in this case.

Finally, it is important to consider which factors did not significantly influence confidence levels. While students' field of study was a significant predictor of performance, it also played a role in confidence levels, with students in physics and engineering showing higher confidence than those in chemistry. However, academic year did not exhibit a strong effect on confidence in this context. This suggests that while disciplinary background contributes to confidence, general academic experience does not necessarily lead to greater self-assurance in conceptual physics questions. This reinforces the idea that confidence in incorrect responses is not merely a function of accumulated coursework but rather stems from deeper cognitive structures and alternative conceptions.

\subsection{How do students' explanation patterns relate to answer accuracy and confidence?}

The analysis of the student's justifications revealed three distinct reasoning patterns:  
(1) \textit{Correct reasoning}, based on differentiation or dimensional analysis;  
(2) \textit{Incorrect but structured reasoning}, where the students argued that the trigonometric derivatives are valid in both degrees and radians;  
(3) \textit{Incoherent or authority-based reasoning}, lacking conceptual justification.  

The most common response (25\%) was \textit{incorrect but well-articulated reasoning}, in which students claimed that trigonometric derivatives are equally valid regardless of angular units. This structured yet incorrect reasoning represents a cognitive ``blind spot,'' where students do not recognize the need for consistency between angular units and differentiation rules.  

Additionally, 50\% of students provided explanations that lacked conceptual depth or relied on authority (e.g., ``this is how it was taught''), while 10\% explicitly expressed uncertainty regarding their answer.  

These findings suggest that the root cause of this misconception is a combination of a deeply ingrained alternative conception alongside a lack of formal knowledge. Together, these factors contribute to students' difficulty in distinguishing between valid and invalid applications of trigonometric derivatives.

\subsection{How does an academic discipline influence students' responses}

The results suggest that academic discipline plays a significant role in students' success in answering questions related to trigonometric derivatives and harmonic oscillators. Engineering students demonstrated the highest success rates, followed by students in mathematics and computer science. In contrast, physics and chemistry students exhibited similar performance levels, despite the mathematical rigor typically associated with physics.

Chi-square analysis indicates that engineering students were the most successful group, followed closely by students in mathematics and computer science. Physics students, however, did not show a clear advantage over chemistry students, suggesting that their mathematical training does not necessarily translate into improved performance in this specific context.

Logistic regression analysis provides additional information by controlling for other variables. The results indicate that engineering students were significantly more likely to answer correctly compared to chemistry students in the trigonometric derivative and harmonic oscillator questions. Mathematics and computer science students demonstrated a moderate advantage, though not always statistically significant, while physics students did not exhibit a measurable advantage over chemistry students.

Interestingly, an analysis of confidence levels provides a different perspective. Although engineering and physics students expressed greater confidence in their responses, their actual success rates differed. This suggests that engineering students’ confidence may be supported by their superior performance, while physics students may feel confident due to their familiarity with the subject rather than explicit proficiency in the mathematical formalism.

These findings may reflect differences in how disciplines approach mathematical reasoning in physics-related contexts. Engineering curricula often emphasize structured mathematical problem-solving and applied differential equations, which could contribute to stronger performance in these tasks. Physics education, by contrast, frequently prioritizes conceptual reasoning and physical intuition over explicit mathematical formalism, which may explain why physics students did not outperform chemistry students in this domain.

A possible factor contributing to the higher success rates of engineering students compared to physics students may be the demanding nature of admissions to engineering programs. In many countries, including India, the United States, and Israel, admission to engineering studies is significantly more competitive compared to the physics and chemistry departments. As a result, engineering students in our sample may have undergone a more rigorous selection process, which could explain their higher performance in this study.

Further research is needed to explore how curricular differences influence students’ ability to apply formal mathematical reasoning in physics-related problems. A comparative study across disciplines, focusing on how mathematical concepts are integrated into physics education, could provide deeper insights into the extent to which instructional approaches shape both conceptual understanding and computational proficiency.

\subsection{How well do grades in foundational courses predict students' answers?}

A key assumption in physics education is that success in foundational courses such as Mechanics and Calculus should correlate with the student's ability to apply fundamental mathematical tools correctly in physical contexts. To test this assumption, we examined whether the students' grades in these courses predicted their success in identifying the conditions under which trigonometric derivatives are valid (Q3) and in correctly applying these derivatives to a physical system (Q4).  

Our findings indicate that this assumption does not hold: self-reported grades in Mechanics and Calculus were not significant predictors of success in either Q3 or Q4. Chi-square analyses revealed no significant associations between performance in these courses and correctness, except for two isolated cases—Calculus scores in Israel for Q3 and Mechanics scores in India for Q4—neither of which held consistently across other countries.  

To further investigate, we conducted a multivariate logistic regression analysis that holds other explanatory variables constant, allowing for a clearer assessment of the independent effect of grades. This approach accounts for potential confounding factors, such as field of study or country of origin, that might otherwise obscure the relationship between grades and performance. In this analysis, neither the mechanics nor the calculus scores were statistically significant predictors of success in Q3 or Q4. The only academic performance metric that showed a significant effect was a high overall GPA (95+), which predicted success in Q4 but not in Q3.  

This result suggests that even students with strong technical backgrounds in Mechanics and Calculus struggle to correctly identify and apply trigonometric derivatives, reinforcing the cognitive blind spot hypothesis. The ability to manipulate mathematical expressions in a computational setting does not necessarily translate to a conceptual understanding of their underlying constraints and applicability in physical scenarios. This disconnect is particularly evident in Q3, where even students with the highest performance in foundational mathematics and physics courses did not demonstrate a higher likelihood of success. These findings align with previous research in Physics Education Research (PER), including Mazur \cite{eric1997peer}, who demonstrated that algorithmic expertise does not ensure conceptual understanding.

\section{Conclusion}

This study reveals a blind spot in the students' understanding of trigonometric derivatives. Surprisingly, the difficulty in distinguishing between radians and degrees is consistent across four different countries and various educational systems, indicating a fundamental challenge in mathematics and physics education. While most students struggle to identify the necessity of radians in pure mathematical contexts, they perform better when encountering the same concept in concrete physical situations. This finding of a physical anchor effect suggests that concrete physical contexts can serve as a cognitive bridge or touchstone) for abstract mathematical concepts \cite{redish2003teaching} . Nevertheless, the fact that approximately \(40\%\) of students err even in physical contexts points to an urgent need to modify how this topic is taught in physics courses.

These findings have significant implications for physics and mathematics instruction. We recommend that calculus and introductory physics courses incorporate dedicated instruction time for trigonometric derivatives, emphasizing the fundamental principles underlying their distinct behavior in radians versus degrees. This approach should integrate multiple evidence-based, interactive pedagogical strategies, e.g.: visual demonstrations comparing trigonometric functions plotted in both units on the same graph, dynamic simulations illustrating their different rates of change, and structured exercises comparing derivatives at various angles. Given students' extensive prior experience with degrees in early education, the transition to radians in higher education requires explicit guidance and clear demonstration of its necessity. At the same time, this approach can build on Arons' notion of "concept first name after" -- beginning with conceptual understanding and moving to more formal representations subsequently \cite{arons1996teaching}. Arons' teaching methods should extend beyond computational success to include assessment tools that evaluate students' conceptual understanding, requiring them to articulate why radians are necessary rather than merely performing mechanical calculations. The integration of physical examples and contexts in both mathematical instruction and practice, starting from the early stages of teaching derivatives, will enable students to recognize inherently where and why radians are essential.

This study has several limitations to consider. Our sample, drawn exclusively from leading research universities, may represent students who are academically better prepared than average. Additionally, while the study included four major countries, there may be other countries where students' understanding patterns of this topic differ, particularly in nations with substantially different educational systems or curricula from those examined. Furthermore, the interpretation of certain findings, such as confidence levels in responses, can be culturally dependent, and cross-cultural comparisons should be approached with caution when dealing with subjective measures collected in different languages. These limitations not only contextualize our findings, but also point to promising directions for future investigation.

Future research could focus on developing and testing educational interventions that emphasize the connection between trigonometric derivatives and the use of radians. Specifically, studies might examine the effectiveness of incorporating visual demonstrations of rates of change in teaching this topic. Furthermore, extending this research to other countries could provide deeper insight into the universality of this phenomenon and identify successful teaching approaches in different educational systems. Moreover, it would be valuable to explore whether the physical anchor effect can be harnessed as a pedagogical tool, specifically whether instruction beginning with physical examples can enhance students' mathematical understanding in more abstract domains as well, following Arons' suggestion \cite{arons1996teaching}.

This study reveals a widespread conceptual failure in understanding trigonometric derivatives: most students do not recognize that these derivatives are valid only in radians, although they identify this better in physical contexts. This finding points to a significant gap between mathematics and physics instruction that requires changes in teaching methods. Without such a change, students will continue to memorize formulas without understanding their deeper meaning.

\section*{Acknowledgements}
The authors thank the students who participated in this study. We also appreciate the contributions of Wu Yuxiao, Roni Segal, Oz Arie, and Einat Abramovitch for their assistance in coding and analyzing the open-ended responses. Additionally, we thank the Physics Education Research (PER) group at the University of Colorado Boulder, in particular Steven J. Pollock and Shams El-Adawy, for their valuable feedback. Special thanks to David A. Kessler for insightful discussions on the topic.

\clearpage

\appendix
\section{Additional Tables and Data}

\begin{table} [ht]
\caption{\label{tab:3} The distribution of responses for the correct case code for trigonometric derivatives by country. Percentages and absolute counts are provided for each response category.}
\begin{ruledtabular}
\begin{tabular}{lcccc}
 & Radians & Both & Degrees & Total \\
\hline
Israel & 31.3\% (85) & 63.2\% (172) & 5.5\% (15) & (272) \\
USA & 18.3\% (30) & 81.7\% (134) & 0.0\% (0) & (164) \\
China & 29.6\% (55) & 68.8\% (128) & 1.6\% (3) & (186) \\
India & 21.8\% (32) & 74.8\% (110) & 3.4\% (5) & (147) \\
Total & 26.3\% (202) & 70.7\% (544) & 3.0\% (23) & (769) \\
\end{tabular}
\end{ruledtabular}
\end{table}

\begin{table}[ht]
\caption{\label{tab:4} The distribution of responses for the correct case code for harmonic oscillator expression by country. Percentages and absolute counts are provided for each response category.}
\begin{ruledtabular}
\begin{tabular}{lcccc}
 & Radians & Both & Degrees & Total \\
\hline
Israel & 62.9\% (171) & 30.5\% (83) & 6.6\% (18) & (272) \\ 
USA & 56.1\% (92) & 36.0\% (59) & 7.9\% (13) & (164) \\
China & 59.7\% (111) & 37.1\% (69) & 3.2\% (6) & (186) \\
India & 54.4\% (80) & 38.1\% (56) & 7.5\% (11) & (147) \\
Total & 59.0\% (454) & 34.7\% (267) & 6.2\% (48) & (769) \\
\end{tabular}
\end{ruledtabular}
\end{table}

\begin{table}[ht]
\caption{\label{tab:33} The distribution of responses for the correct case code for trigonometric derivatives by country for 1-2 years in Physics only. Percentages and absolute counts are provided for each response category.}
\begin{ruledtabular}
\begin{tabular}{lcccc}
 & Radians & Both & Degrees & Total \\
\hline
Israel & 20.2\% (8) & 80.0\% (32) & 0.0\% (0) & (40) \\
USA & 16.5\% (14) & 83.5\% (71) & 0.0\% (0) & (85) \\
China & 27.3\% (9) & 69.7\% (23) & 3.0\% (1) & (33) \\
Total & 19.6\% (31) & 79.7\% (126) & 0.6\% (1) & (158) \\
\end{tabular}
\end{ruledtabular}
\end{table}

\begin{table}[ht]
\caption{\label{tab:34} The distribution of responses for the correct case code for Harmonic Oscillator derivatives by country for 1-2 years in Physics only. Percentages and absolute counts are provided for each response category.}
\begin{ruledtabular}
\begin{tabular}{lcccc}
 & Radians & Both & Degrees & Total \\
\hline
Israel & 55.0\% (22) & 37.5\% (15) & 7.5\% (3) & (40) \\
USA & 55.3\% (47) & 37.6\% (32) & 7.1\% (6) & (85) \\
China & 57.6\% (19) & 42.4\% (14) & 0.0\% (0) & (33) \\
Total & 55.7\% (88) & 38.6\% (61) & 5.7\% (9) & (158) \\
\end{tabular}
\end{ruledtabular}
\end{table}

\begin{table} [ht]
\caption{\label{tab:5}
The distribution of respondent types by country}
\begin{ruledtabular}
\begin{tabular}{lccccccc}
Country & Consistent  &  Correct  &  Correct  & Consistent  & Total \\
  &  Correct & Q3 Only & Q4 Only &  Incorrect &  \\
\hline
Israel & 26.1\%  & 5.1\%  & 36.8\%  & 32.0\%  &  \\
 & (71) & (14) & (100) & (87) & (272)  \\
USA & 15.2\%  & 3.0\%  & 40.9\% & 40.9\%  &  \\
 & (25) & (5)  & (67) & (67) & (164)  \\
China & 21.0\%  & 8.6\%  & 38.7\%  & 31.7\%  &  \\
 & (39) & (16) & (72) & (59) & (186)  \\
India & 15.0\% & 6.8\%  & 39.5\%  & 38.8\%  &  \\
 & (22)  & (10) & (58) & (57) & (147)  \\
Total & 20.4\%  & 5.9\%  & 38.6\%  & 35.1\%  &  \\
 & (157) & (45) & (297) & (270) & (769)  \\
\end{tabular}
\end{ruledtabular}
\end{table}

\begin{table}[ht]
\caption{\label{tab:7}
The distribution of respondent types by country}
\begin{ruledtabular}
\begin{tabular}{lcccc}
Country & Meaningful & Misguided & Meaningless & Total \\
 & Reasoning & Reasoning & Reasoning &  \\ 
 \hline
Israel & 21.3\% (58) & 18.4\% (50) & 60.3\% (164) & (272) \\
USA & 26.2\% (43) & 32.3\% (53) & 41.5\% (68) & (164) \\
China & 20.4\% (38) & 24.2\% (45) & 55.4\% (103) & (186) \\
India & 27.9\% (41) & 24.5\% (36) & 47.6\% (70) & (147) \\
Total & 23.4\% (180) & 23.9\% (184) & 52.7\% (405) & (769) \\
\end{tabular}
\end{ruledtabular}
\end{table}

\clearpage

\bibliography{apssamp}

\begin{thebibliography}{28}%
\makeatletter
\providecommand \@ifxundefined [1]{%
 \@ifx{#1\undefined}
}%
\providecommand \@ifnum [1]{%
 \ifnum #1\expandafter \@firstoftwo
 \else \expandafter \@secondoftwo
 \fi
}%
\providecommand \@ifx [1]{%
 \ifx #1\expandafter \@firstoftwo
 \else \expandafter \@secondoftwo
 \fi
}%
\providecommand \natexlab [1]{#1}%
\providecommand \enquote  [1]{``#1''}%
\providecommand \bibnamefont  [1]{#1}%
\providecommand \bibfnamefont [1]{#1}%
\providecommand \citenamefont [1]{#1}%
\providecommand \href@noop [0]{\@secondoftwo}%
\providecommand \href [0]{\begingroup \@sanitize@url \@href}%
\providecommand \@href[1]{\@@startlink{#1}\@@href}%
\providecommand \@@href[1]{\endgroup#1\@@endlink}%
\providecommand \@sanitize@url [0]{\catcode `\\12\catcode `\$12\catcode `\&12\catcode `\#12\catcode `\^12\catcode `\_12\catcode `\%12\relax}%
\providecommand \@@startlink[1]{}%
\providecommand \@@endlink[0]{}%
\providecommand \url  [0]{\begingroup\@sanitize@url \@url }%
\providecommand \@url [1]{\endgroup\@href {#1}{\urlprefix }}%
\providecommand \urlprefix  [0]{URL }%
\providecommand \Eprint [0]{\href }%
\providecommand \doibase [0]{https://doi.org/}%
\providecommand \selectlanguage [0]{\@gobble}%
\providecommand \bibinfo  [0]{\@secondoftwo}%
\providecommand \bibfield  [0]{\@secondoftwo}%
\providecommand \translation [1]{[#1]}%
\providecommand \BibitemOpen [0]{}%
\providecommand \bibitemStop [0]{}%
\providecommand \bibitemNoStop [0]{.\EOS\space}%
\providecommand \EOS [0]{\spacefactor3000\relax}%
\providecommand \BibitemShut  [1]{\csname bibitem#1\endcsname}%
\let\auto@bib@innerbib\@empty
\bibitem [{\citenamefont {Shankar}(2019)}]{shankar2019fundamentals}%
  \BibitemOpen
  \bibfield  {author} {\bibinfo {author} {\bibfnamefont {R.}~\bibnamefont {Shankar}},\ }\href@noop {} {\emph {\bibinfo {title} {Fundamentals of physics I: mechanics, relativity, and thermodynamics}}}\ (\bibinfo  {publisher} {Yale University Press},\ \bibinfo {year} {2019})\BibitemShut {NoStop}%
\bibitem [{\citenamefont {Bollen}\ \emph {et~al.}(2016)\citenamefont {Bollen}, \citenamefont {van Kampen}, \citenamefont {Baily},\ and\ \citenamefont {De~Cock}}]{PhysRevPhysEducRes.12.020134}%
  \BibitemOpen
  \bibfield  {author} {\bibinfo {author} {\bibfnamefont {L.}~\bibnamefont {Bollen}}, \bibinfo {author} {\bibfnamefont {P.}~\bibnamefont {van Kampen}}, \bibinfo {author} {\bibfnamefont {C.}~\bibnamefont {Baily}},\ and\ \bibinfo {author} {\bibfnamefont {M.}~\bibnamefont {De~Cock}},\ }\bibfield  {title} {\bibinfo {title} {Qualitative investigation into students' use of divergence and curl in electromagnetism},\ }\href {https://doi.org/10.1103/PhysRevPhysEducRes.12.020134} {\bibfield  {journal} {\bibinfo  {journal} {Phys. Rev. Phys. Educ. Res.}\ }\textbf {\bibinfo {volume} {12}},\ \bibinfo {pages} {020134} (\bibinfo {year} {2016})}\BibitemShut {NoStop}%
\bibitem [{\citenamefont {Yeatts}\ and\ \citenamefont {Hundhausen}(1992)}]{10.1119/1.17077}%
  \BibitemOpen
  \bibfield  {author} {\bibinfo {author} {\bibfnamefont {F.~R.}\ \bibnamefont {Yeatts}}\ and\ \bibinfo {author} {\bibfnamefont {J.~R.}\ \bibnamefont {Hundhausen}},\ }\bibfield  {title} {\bibinfo {title} {Calculus and physics: Challenges at the interface},\ }\href {https://doi.org/10.1119/1.17077} {\bibfield  {journal} {\bibinfo  {journal} {American Journal of Physics}\ }\textbf {\bibinfo {volume} {60}},\ \bibinfo {pages} {716} (\bibinfo {year} {1992})}\BibitemShut {NoStop}%
\bibitem [{\citenamefont {Wilcox}\ and\ \citenamefont {Pollock}(2015{\natexlab{a}})}]{PhysRevSTPER.11.010108}%
  \BibitemOpen
  \bibfield  {author} {\bibinfo {author} {\bibfnamefont {B.~R.}\ \bibnamefont {Wilcox}}\ and\ \bibinfo {author} {\bibfnamefont {S.~J.}\ \bibnamefont {Pollock}},\ }\bibfield  {title} {\bibinfo {title} {Upper-division student difficulties with the dirac delta function},\ }\href {https://doi.org/10.1103/PhysRevSTPER.11.010108} {\bibfield  {journal} {\bibinfo  {journal} {Phys. Rev. ST Phys. Educ. Res.}\ }\textbf {\bibinfo {volume} {11}},\ \bibinfo {pages} {010108} (\bibinfo {year} {2015}{\natexlab{a}})}\BibitemShut {NoStop}%
\bibitem [{\citenamefont {Bollen}\ \emph {et~al.}(2015)\citenamefont {Bollen}, \citenamefont {van Kampen},\ and\ \citenamefont {De~Cock}}]{PhysRevSTPER.11.020129}%
  \BibitemOpen
  \bibfield  {author} {\bibinfo {author} {\bibfnamefont {L.}~\bibnamefont {Bollen}}, \bibinfo {author} {\bibfnamefont {P.}~\bibnamefont {van Kampen}},\ and\ \bibinfo {author} {\bibfnamefont {M.}~\bibnamefont {De~Cock}},\ }\bibfield  {title} {\bibinfo {title} {Students' difficulties with vector calculus in electrodynamics},\ }\href {https://doi.org/10.1103/PhysRevSTPER.11.020129} {\bibfield  {journal} {\bibinfo  {journal} {Phys. Rev. ST Phys. Educ. Res.}\ }\textbf {\bibinfo {volume} {11}},\ \bibinfo {pages} {020129} (\bibinfo {year} {2015})}\BibitemShut {NoStop}%
\bibitem [{\citenamefont {Wilcox}\ and\ \citenamefont {Pollock}(2015{\natexlab{b}})}]{PhysRevSTPER.11.020131}%
  \BibitemOpen
  \bibfield  {author} {\bibinfo {author} {\bibfnamefont {B.~R.}\ \bibnamefont {Wilcox}}\ and\ \bibinfo {author} {\bibfnamefont {S.~J.}\ \bibnamefont {Pollock}},\ }\bibfield  {title} {\bibinfo {title} {Upper-division student difficulties with separation of variables},\ }\href {https://doi.org/10.1103/PhysRevSTPER.11.020131} {\bibfield  {journal} {\bibinfo  {journal} {Phys. Rev. ST Phys. Educ. Res.}\ }\textbf {\bibinfo {volume} {11}},\ \bibinfo {pages} {020131} (\bibinfo {year} {2015}{\natexlab{b}})}\BibitemShut {NoStop}%
\bibitem [{\citenamefont {Nguyen}\ and\ \citenamefont {Rebello}(2011)}]{PhysRevSTPER.7.010113}%
  \BibitemOpen
  \bibfield  {author} {\bibinfo {author} {\bibfnamefont {D.-H.}\ \bibnamefont {Nguyen}}\ and\ \bibinfo {author} {\bibfnamefont {N.~S.}\ \bibnamefont {Rebello}},\ }\bibfield  {title} {\bibinfo {title} {Students' difficulties with integration in electricity},\ }\href {https://doi.org/10.1103/PhysRevSTPER.7.010113} {\bibfield  {journal} {\bibinfo  {journal} {Phys. Rev. ST Phys. Educ. Res.}\ }\textbf {\bibinfo {volume} {7}},\ \bibinfo {pages} {010113} (\bibinfo {year} {2011})}\BibitemShut {NoStop}%
\bibitem [{\citenamefont {Pepper}\ \emph {et~al.}(2012)\citenamefont {Pepper}, \citenamefont {Chasteen}, \citenamefont {Pollock},\ and\ \citenamefont {Perkins}}]{PhysRevSTPER.8.010111}%
  \BibitemOpen
  \bibfield  {author} {\bibinfo {author} {\bibfnamefont {R.~E.}\ \bibnamefont {Pepper}}, \bibinfo {author} {\bibfnamefont {S.~V.}\ \bibnamefont {Chasteen}}, \bibinfo {author} {\bibfnamefont {S.~J.}\ \bibnamefont {Pollock}},\ and\ \bibinfo {author} {\bibfnamefont {K.~K.}\ \bibnamefont {Perkins}},\ }\bibfield  {title} {\bibinfo {title} {Observations on student difficulties with mathematics in upper-division electricity and magnetism},\ }\href {https://doi.org/10.1103/PhysRevSTPER.8.010111} {\bibfield  {journal} {\bibinfo  {journal} {Phys. Rev. ST Phys. Educ. Res.}\ }\textbf {\bibinfo {volume} {8}},\ \bibinfo {pages} {010111} (\bibinfo {year} {2012})}\BibitemShut {NoStop}%
\bibitem [{\citenamefont {Cui}\ \emph {et~al.}(2006)\citenamefont {Cui}, \citenamefont {Rebello},\ and\ \citenamefont {Bennett}}]{10.1063/1.2177017}%
  \BibitemOpen
  \bibfield  {author} {\bibinfo {author} {\bibfnamefont {L.}~\bibnamefont {Cui}}, \bibinfo {author} {\bibfnamefont {N.~S.}\ \bibnamefont {Rebello}},\ and\ \bibinfo {author} {\bibfnamefont {A.~G.}\ \bibnamefont {Bennett}},\ }\bibfield  {title} {\bibinfo {title} {College students’ transfer from calculus to physics},\ }\href {https://doi.org/10.1063/1.2177017} {\bibfield  {journal} {\bibinfo  {journal} {AIP Conference Proceedings}\ }\textbf {\bibinfo {volume} {818}},\ \bibinfo {pages} {37} (\bibinfo {year} {2006})}\BibitemShut {NoStop}%
\bibitem [{\citenamefont {Carli}\ \emph {et~al.}(2020)\citenamefont {Carli}, \citenamefont {Lippiello}, \citenamefont {Pantano}, \citenamefont {Perona},\ and\ \citenamefont {Tormen}}]{PhysRevPhysEducRes.16.010111}%
  \BibitemOpen
  \bibfield  {author} {\bibinfo {author} {\bibfnamefont {M.}~\bibnamefont {Carli}}, \bibinfo {author} {\bibfnamefont {S.}~\bibnamefont {Lippiello}}, \bibinfo {author} {\bibfnamefont {O.}~\bibnamefont {Pantano}}, \bibinfo {author} {\bibfnamefont {M.}~\bibnamefont {Perona}},\ and\ \bibinfo {author} {\bibfnamefont {G.}~\bibnamefont {Tormen}},\ }\bibfield  {title} {\bibinfo {title} {Testing students ability to use derivatives, integrals, and vectors in a purely mathematical context and in a physical context},\ }\href {https://doi.org/10.1103/PhysRevPhysEducRes.16.010111} {\bibfield  {journal} {\bibinfo  {journal} {Phys. Rev. Phys. Educ. Res.}\ }\textbf {\bibinfo {volume} {16}},\ \bibinfo {pages} {010111} (\bibinfo {year} {2020})}\BibitemShut {NoStop}%
\bibitem [{\citenamefont {Gifford}\ and\ \citenamefont {Finkelstein}(2020)}]{PhysRevPhysEducRes.16.020121}%
  \BibitemOpen
  \bibfield  {author} {\bibinfo {author} {\bibfnamefont {J.~D.}\ \bibnamefont {Gifford}}\ and\ \bibinfo {author} {\bibfnamefont {N.~D.}\ \bibnamefont {Finkelstein}},\ }\bibfield  {title} {\bibinfo {title} {Categorical framework for mathematical sense making in physics},\ }\href {https://doi.org/10.1103/PhysRevPhysEducRes.16.020121} {\bibfield  {journal} {\bibinfo  {journal} {Phys. Rev. Phys. Educ. Res.}\ }\textbf {\bibinfo {volume} {16}},\ \bibinfo {pages} {020121} (\bibinfo {year} {2020})}\BibitemShut {NoStop}%
\bibitem [{\citenamefont {Gifford}\ and\ \citenamefont {Finkelstein}(2021)}]{PhysRevPhysEducRes.17.010138}%
  \BibitemOpen
  \bibfield  {author} {\bibinfo {author} {\bibfnamefont {J.~D.}\ \bibnamefont {Gifford}}\ and\ \bibinfo {author} {\bibfnamefont {N.~D.}\ \bibnamefont {Finkelstein}},\ }\bibfield  {title} {\bibinfo {title} {Applying a mathematical sense-making framework to student work and its potential for curriculum design},\ }\href {https://doi.org/10.1103/PhysRevPhysEducRes.17.010138} {\bibfield  {journal} {\bibinfo  {journal} {Phys. Rev. Phys. Educ. Res.}\ }\textbf {\bibinfo {volume} {17}},\ \bibinfo {pages} {010138} (\bibinfo {year} {2021})}\BibitemShut {NoStop}%
\bibitem [{\citenamefont {Galle}\ and\ \citenamefont {Meredith}(2014)}]{10.1119/1.4862119}%
  \BibitemOpen
  \bibfield  {author} {\bibinfo {author} {\bibfnamefont {G.}~\bibnamefont {Galle}}\ and\ \bibinfo {author} {\bibfnamefont {D.}~\bibnamefont {Meredith}},\ }\bibfield  {title} {\bibinfo {title} {The trouble with trig},\ }\href {https://doi.org/10.1119/1.4862119} {\bibfield  {journal} {\bibinfo  {journal} {The Physics Teacher}\ }\textbf {\bibinfo {volume} {52}},\ \bibinfo {pages} {112} (\bibinfo {year} {2014})}\BibitemShut {NoStop}%
\bibitem [{\citenamefont {Weber}(2005)}]{weber2005students}%
  \BibitemOpen
  \bibfield  {author} {\bibinfo {author} {\bibfnamefont {K.}~\bibnamefont {Weber}},\ }\bibfield  {title} {\bibinfo {title} {Students’ understanding of trigonometric functions},\ }\href@noop {} {\bibfield  {journal} {\bibinfo  {journal} {Mathematics Education Research Journal}\ }\textbf {\bibinfo {volume} {17}},\ \bibinfo {pages} {91} (\bibinfo {year} {2005})}\BibitemShut {NoStop}%
\bibitem [{\citenamefont {Halliday}\ \emph {et~al.}(2013)\citenamefont {Halliday}, \citenamefont {Resnick},\ and\ \citenamefont {Walker}}]{halliday2013fundamentals}%
  \BibitemOpen
  \bibfield  {author} {\bibinfo {author} {\bibfnamefont {D.}~\bibnamefont {Halliday}}, \bibinfo {author} {\bibfnamefont {R.}~\bibnamefont {Resnick}},\ and\ \bibinfo {author} {\bibfnamefont {J.}~\bibnamefont {Walker}},\ }\href@noop {} {\emph {\bibinfo {title} {Fundamentals of physics}}}\ (\bibinfo  {publisher} {John Wiley \& Sons},\ \bibinfo {year} {2013})\BibitemShut {NoStop}%
\bibitem [{\citenamefont {Serway}\ \emph {et~al.}(2000)\citenamefont {Serway}, \citenamefont {Jewett},\ and\ \citenamefont {Peroomian}}]{serway2000physics}%
  \BibitemOpen
  \bibfield  {author} {\bibinfo {author} {\bibfnamefont {R.~A.}\ \bibnamefont {Serway}}, \bibinfo {author} {\bibfnamefont {J.~W.}\ \bibnamefont {Jewett}},\ and\ \bibinfo {author} {\bibfnamefont {V.}~\bibnamefont {Peroomian}},\ }\href@noop {} {\emph {\bibinfo {title} {Physics for scientists and engineers}}},\ Vol.~\bibinfo {volume} {2}\ (\bibinfo  {publisher} {Saunders college publishing Philadelphia},\ \bibinfo {year} {2000})\BibitemShut {NoStop}%
\bibitem [{\citenamefont {Tipler}\ and\ \citenamefont {Mosca}(2007)}]{tipler2007physics}%
  \BibitemOpen
  \bibfield  {author} {\bibinfo {author} {\bibfnamefont {P.~A.}\ \bibnamefont {Tipler}}\ and\ \bibinfo {author} {\bibfnamefont {G.}~\bibnamefont {Mosca}},\ }\href@noop {} {\emph {\bibinfo {title} {Physics for scientists and engineers}}}\ (\bibinfo  {publisher} {Macmillan},\ \bibinfo {year} {2007})\BibitemShut {NoStop}%
\bibitem [{\citenamefont {Beichner}(1994)}]{beichner1994testing}%
  \BibitemOpen
  \bibfield  {author} {\bibinfo {author} {\bibfnamefont {R.~J.}\ \bibnamefont {Beichner}},\ }\bibfield  {title} {\bibinfo {title} {Testing student interpretation of kinematics graphs},\ }\href@noop {} {\bibfield  {journal} {\bibinfo  {journal} {American journal of Physics}\ }\textbf {\bibinfo {volume} {62}},\ \bibinfo {pages} {750} (\bibinfo {year} {1994})}\BibitemShut {NoStop}%
\bibitem [{\citenamefont {Christensen}\ and\ \citenamefont {Thompson}(2012)}]{christensen2012investigating}%
  \BibitemOpen
  \bibfield  {author} {\bibinfo {author} {\bibfnamefont {W.~M.}\ \bibnamefont {Christensen}}\ and\ \bibinfo {author} {\bibfnamefont {J.~R.}\ \bibnamefont {Thompson}},\ }\bibfield  {title} {\bibinfo {title} {Investigating graphical representations of slope and derivative without a physics context},\ }\href@noop {} {\bibfield  {journal} {\bibinfo  {journal} {Physical Review Special Topics—Physics Education Research}\ }\textbf {\bibinfo {volume} {8}},\ \bibinfo {pages} {023101} (\bibinfo {year} {2012})}\BibitemShut {NoStop}%
\bibitem [{\citenamefont {Bao}\ \emph {et~al.}(2009{\natexlab{a}})\citenamefont {Bao}, \citenamefont {Fang}, \citenamefont {Cai}, \citenamefont {Wang}, \citenamefont {Yang}, \citenamefont {Cui}, \citenamefont {Han}, \citenamefont {Ding},\ and\ \citenamefont {Luo}}]{10.1119/1.2976334}%
  \BibitemOpen
  \bibfield  {author} {\bibinfo {author} {\bibfnamefont {L.}~\bibnamefont {Bao}}, \bibinfo {author} {\bibfnamefont {K.}~\bibnamefont {Fang}}, \bibinfo {author} {\bibfnamefont {T.}~\bibnamefont {Cai}}, \bibinfo {author} {\bibfnamefont {J.}~\bibnamefont {Wang}}, \bibinfo {author} {\bibfnamefont {L.}~\bibnamefont {Yang}}, \bibinfo {author} {\bibfnamefont {L.}~\bibnamefont {Cui}}, \bibinfo {author} {\bibfnamefont {J.}~\bibnamefont {Han}}, \bibinfo {author} {\bibfnamefont {L.}~\bibnamefont {Ding}},\ and\ \bibinfo {author} {\bibfnamefont {Y.}~\bibnamefont {Luo}},\ }\bibfield  {title} {\bibinfo {title} {Learning of content knowledge and development of scientific reasoning ability: A cross culture comparison},\ }\href {https://doi.org/10.1119/1.2976334} {\bibfield  {journal} {\bibinfo  {journal} {American Journal of Physics}\ }\textbf {\bibinfo {volume} {77}},\ \bibinfo {pages} {1118} (\bibinfo {year} {2009}{\natexlab{a}})},\ \Eprint
  {https://arxiv.org/abs/https://pubs.aip.org/aapt/ajp/article-pdf/77/12/1118/13093738/1118\_1\_online.pdf} {https://pubs.aip.org/aapt/ajp/article-pdf/77/12/1118/13093738/1118\_1\_online.pdf} \BibitemShut {NoStop}%
\bibitem [{\citenamefont {Bao}\ \emph {et~al.}(2009{\natexlab{b}})\citenamefont {Bao}, \citenamefont {Cai}, \citenamefont {Koenig}, \citenamefont {Fang}, \citenamefont {Han}, \citenamefont {Wang}, \citenamefont {Liu}, \citenamefont {Ding}, \citenamefont {Cui}, \citenamefont {Luo}, \citenamefont {Wang}, \citenamefont {Li},\ and\ \citenamefont {Wu}}]{doi:10.1126/science.1167740}%
  \BibitemOpen
  \bibfield  {author} {\bibinfo {author} {\bibfnamefont {L.}~\bibnamefont {Bao}}, \bibinfo {author} {\bibfnamefont {T.}~\bibnamefont {Cai}}, \bibinfo {author} {\bibfnamefont {K.}~\bibnamefont {Koenig}}, \bibinfo {author} {\bibfnamefont {K.}~\bibnamefont {Fang}}, \bibinfo {author} {\bibfnamefont {J.}~\bibnamefont {Han}}, \bibinfo {author} {\bibfnamefont {J.}~\bibnamefont {Wang}}, \bibinfo {author} {\bibfnamefont {Q.}~\bibnamefont {Liu}}, \bibinfo {author} {\bibfnamefont {L.}~\bibnamefont {Ding}}, \bibinfo {author} {\bibfnamefont {L.}~\bibnamefont {Cui}}, \bibinfo {author} {\bibfnamefont {Y.}~\bibnamefont {Luo}}, \bibinfo {author} {\bibfnamefont {Y.}~\bibnamefont {Wang}}, \bibinfo {author} {\bibfnamefont {L.}~\bibnamefont {Li}},\ and\ \bibinfo {author} {\bibfnamefont {N.}~\bibnamefont {Wu}},\ }\bibfield  {title} {\bibinfo {title} {Learning and scientific reasoning},\ }\href {https://doi.org/10.1126/science.1167740} {\bibfield  {journal} {\bibinfo  {journal} {Science}\ }\textbf {\bibinfo {volume} {323}},\
  \bibinfo {pages} {586} (\bibinfo {year} {2009}{\natexlab{b}})},\ \Eprint {https://arxiv.org/abs/https://www.science.org/doi/pdf/10.1126/science.1167740} {https://www.science.org/doi/pdf/10.1126/science.1167740} \BibitemShut {NoStop}%
\bibitem [{\citenamefont {Hasan}\ \emph {et~al.}(1999)\citenamefont {Hasan}, \citenamefont {Bagayoko},\ and\ \citenamefont {Kelley}}]{Saleem_Hasan_1999}%
  \BibitemOpen
  \bibfield  {author} {\bibinfo {author} {\bibfnamefont {S.}~\bibnamefont {Hasan}}, \bibinfo {author} {\bibfnamefont {D.}~\bibnamefont {Bagayoko}},\ and\ \bibinfo {author} {\bibfnamefont {E.~L.}\ \bibnamefont {Kelley}},\ }\bibfield  {title} {\bibinfo {title} {Misconceptions and the certainty of response index (cri)},\ }\href {https://doi.org/10.1088/0031-9120/34/5/304} {\bibfield  {journal} {\bibinfo  {journal} {Physics Education}\ }\textbf {\bibinfo {volume} {34}},\ \bibinfo {pages} {294} (\bibinfo {year} {1999})}\BibitemShut {NoStop}%
\bibitem [{\citenamefont {Potgieter}\ \emph {et~al.}(2010)\citenamefont {Potgieter}, \citenamefont {Malatje}, \citenamefont {Gaigher},\ and\ \citenamefont {Venter}}]{potgieter2010confidence}%
  \BibitemOpen
  \bibfield  {author} {\bibinfo {author} {\bibfnamefont {M.}~\bibnamefont {Potgieter}}, \bibinfo {author} {\bibfnamefont {E.}~\bibnamefont {Malatje}}, \bibinfo {author} {\bibfnamefont {E.}~\bibnamefont {Gaigher}},\ and\ \bibinfo {author} {\bibfnamefont {E.}~\bibnamefont {Venter}},\ }\bibfield  {title} {\bibinfo {title} {Confidence versus performance as an indicator of the presence of alternative conceptions and inadequate problem-solving skills in mechanics},\ }\href@noop {} {\bibfield  {journal} {\bibinfo  {journal} {International journal of science education}\ }\textbf {\bibinfo {volume} {32}},\ \bibinfo {pages} {1407} (\bibinfo {year} {2010})}\BibitemShut {NoStop}%
\bibitem [{\citenamefont {Wright}\ and\ \citenamefont {Wolff}(2024)}]{wright2024justifying}%
  \BibitemOpen
  \bibfield  {author} {\bibinfo {author} {\bibfnamefont {D.~B.}\ \bibnamefont {Wright}}\ and\ \bibinfo {author} {\bibfnamefont {S.~M.}\ \bibnamefont {Wolff}},\ }\bibfield  {title} {\bibinfo {title} {Justifying responses affects the relationship between confidence and accuracy.},\ }\href@noop {} {\bibfield  {journal} {\bibinfo  {journal} {Experimental Psychology}\ } (\bibinfo {year} {2024})}\BibitemShut {NoStop}%
\bibitem [{\citenamefont {Stankov}\ and\ \citenamefont {Lee}(2014)}]{stankov2014overconfidence}%
  \BibitemOpen
  \bibfield  {author} {\bibinfo {author} {\bibfnamefont {L.}~\bibnamefont {Stankov}}\ and\ \bibinfo {author} {\bibfnamefont {J.}~\bibnamefont {Lee}},\ }\bibfield  {title} {\bibinfo {title} {Overconfidence across world regions},\ }\href@noop {} {\bibfield  {journal} {\bibinfo  {journal} {Journal of Cross-Cultural Psychology}\ }\textbf {\bibinfo {volume} {45}},\ \bibinfo {pages} {821} (\bibinfo {year} {2014})}\BibitemShut {NoStop}%
\bibitem [{\citenamefont {Eric}(1997)}]{eric1997peer}%
  \BibitemOpen
  \bibfield  {author} {\bibinfo {author} {\bibfnamefont {M.}~\bibnamefont {Eric}},\ }\href@noop {} {\bibinfo {title} {Peer instruction: A user’s manual}} (\bibinfo {year} {1997})\BibitemShut {NoStop}%
\bibitem [{\citenamefont {Redish}\ and\ \citenamefont {Burciaga}(2003)}]{redish2003teaching}%
  \BibitemOpen
  \bibfield  {author} {\bibinfo {author} {\bibfnamefont {E.~F.}\ \bibnamefont {Redish}}\ and\ \bibinfo {author} {\bibfnamefont {J.~R.}\ \bibnamefont {Burciaga}},\ }\href@noop {} {\emph {\bibinfo {title} {Teaching physics: With the physics suite}}},\ Vol.~\bibinfo {volume} {1}\ (\bibinfo  {publisher} {John Wiley \& Sons Hoboken, NJ},\ \bibinfo {year} {2003})\BibitemShut {NoStop}%
\bibitem [{\citenamefont {Arons}(1996)}]{arons1996teaching}%
  \BibitemOpen
  \bibfield  {author} {\bibinfo {author} {\bibfnamefont {A.~B.}\ \bibnamefont {Arons}},\ }\href@noop {} {\emph {\bibinfo {title} {Teaching introductory physics}}}\ (\bibinfo  {publisher} {Wiley, New York},\ \bibinfo {year} {1996})\BibitemShut {NoStop}%
\end{thebibliography}%

\end{document}